\documentclass[11pt, letterpaper]{article}
\usepackage{graphicx}
\usepackage{amsmath}
\usepackage{amssymb}
\usepackage{setspace}
\usepackage{epstopdf}
\usepackage{color}
\usepackage[letterpaper,left=1in,right=1in,top=1in,bottom=1in]{geometry}
\usepackage{multirow}
\usepackage{longtable}
\usepackage{rotating}
\usepackage{setspace}
\usepackage{geometry}
\usepackage{subcaption}
\usepackage{amsmath}
\usepackage{algorithm}
\usepackage{algpseudocode}
\usepackage{mathrsfs}
\usepackage{lineno}

\algnewcommand\Initialization{\item[\textbf{Initialization:}]}%

\usepackage{pgfplots}
\pgfplotsset{compat=1.9}
\usetikzlibrary{plotmarks}
\usepackage{float}

\renewcommand{\arraystretch}{0.6}

\title{\Large \bf Performance triggered adaptive model reduction for soil moisture estimation in precision irrigation}

\author{
 \centerline{\normalsize Sarupa Debnath$^{a}$, Bernard Twum Agyeman$^{a}$, Soumya Ranjan Sahoo$^{a}$, Xunyuan Yin$^{b}$,}\\
 \centerline{\normalsize Jinfeng Liu$^{a,}$\thanks{Corresponding author: J. Liu. Tel: +1-780-492-1317. Fax: +1-780-492-2881. Email: jinfeng@ualberta.ca.}}
\vspace{3mm}\\
\centerline{\small $^{a}$Department of Chemical \& Materials Engineering, University of Alberta,}\\
\centerline{\small Edmonton, AB T6G 1H9, Canada}\\
\centerline{\small $^{b}$School of Chemistry, Chemical Engineering \& Biotechnology, }\\
\centerline{\small Nanyang Technological University, 62 Nanyang Drive, Singapore, 637459, Singapore}
}
\allowdisplaybreaks
\begin{document}
\date{}
\maketitle
\setstretch{1.5}
\begin{abstract}
Accurate soil moisture information is crucial for developing precise irrigation control strategies to enhance water use efficiency. Soil moisture estimation based on limited soil moisture sensors is crucial for obtaining comprehensive soil moisture information when dealing with large-scale agricultural fields. The major challenge in soil moisture estimation lies in the high dimensionality of the spatially discretized agro-hydrological models. In this work, we propose a performance-triggered adaptive model reduction approach to address this challenge. The proposed approach employs a trajectory-based unsupervised machine learning technique, and a prediction performance-based triggering scheme is designed to govern model updates adaptively in a way such that the prediction error between the reduced model and the original model over a prediction horizon is maintained below a predetermined threshold. An adaptive extended Kalman filter (EKF) is designed based on the reduced model for soil moisture estimation. The applicability and performance of the proposed approach are evaluated extensively through the application to a simulated large-scale agricultural field. 
\end{abstract}

\clearpage

\section{Introduction}
With population growth, climate change, and environmental pollution, freshwater scarcity has become a global risk \cite{global_2015}. Agricultural irrigation is the primary consumer of freshwater \cite{wastewater_2017}; however, the water use efficiency in agricultural irrigation is low \cite{lozoya_model_2014}. Improving water use efficiency in agricultural irrigation through precision irrigation is a critical step in managing the water crisis. Closed-loop irrigation based on real-time field soil moisture information has been recognized as a promising technical path to realizing precision irrigation \cite{goodchild2015method, mccarthy2014simulation}. Previous studies have shown that closed-loop irrigation based on real-time field soil moisture data can significantly increase water use efficiency \cite{mao2018soil, bwambale2022smart, saavoss2016yield}.

The adoption of closed-loop irrigation systems in greenhouse applications and nursery industries has been on the rise \cite{bwambale2022smart}. Previous studies, including \cite{saavoss2016yield, lichtenberg2013profitability}, have demonstrated that implementing a closed-loop irrigation system in greenhouses, with sensor-monitored control, positively impacts crop yield, quality, and overall profitability. Another study \cite{chappell2013implementation} highlights that real-time soil moisture monitoring, enabled by wireless sensor networks, improves irrigation scheduling in container nurseries, resulting in shorter cropping cycles and expanded production within existing water resources. Notably, these studies focus on greenhouses and small-scale container nurseries. However, when considering the implementation of a sensor-based controller for large-scale agricultural systems, the major challenge lies in obtaining comprehensive field-wise soil moisture information. Unfortunately, acquiring such measurements presents a significant hurdle when dealing with large-scale agricultural fields. The absence of precise soil moisture information often results in irrigation practices that are heavily reliant on farmers' experience and observations rather than on actual field conditions.

State estimation can enable real-time estimation of the soil moisture content across the agricultural field by utilizing the measurements collected from the available sensors. To achieve a comprehensive understanding of the dynamic behaviors of the soil moisture across the entire field, the implementation of an appropriate state estimation method is essential. Several well-known algorithms like the extended Kalman filter (EKF) \cite{lu2011dual}, ensemble Kalman filter (EnKF) \cite{chen2015comparison}, particle filter \cite{montzka2011hydraulic}, and moving horizon estimation (MHE) \cite{bo2020parameter, yin2021consensus} have been applied for soil moisture estimation. These methods have addressed different challenges in the soil moisture estimation of a field. In \cite{agyeman2021soil}, a soil moisture map was constructed by assimilating microwave remote sensor measurements into the 3D Richards equation using the EKF. While in \cite{bo2020decentralized}, state estimation was performed using tensiometer measurements based on the 1D Richards equation. However, estimating soil water content accurately is a major challenge, given the complex high dimensionality of agro-hydrological systems. 

The Richards equation plays an important role in modeling soil water dynamics in agro-hydrological systems \cite{richards_capillary_1931}. To maintain numerical stability and adhere to the local equilibrium in soil water and water potential, a fine discretization (ranging from a few centimeters to several meters) is typically necessary. This results in a large number of nodes which typically ranges from $10^4$ to $10^8$; this poses significant challenges when seeking numerical solutions. For instance, the online optimization associated with nonlinear MHE can become intractable due to the high dimensionality of the system. The EKF is a widely recognized extension of the traditional Kalman filter for handling estimation of nonlinear systems. However, linearizing the nonlinear model consecutively and propagating the covariance matrix explicitly to handle the nonlinear observation is yet another challenging issue for a very large-scale agro-hydrological model obtained from discretizing the Richards equation. 

Model order reduction is a promising framework to reduce the dimensionality of a high-dimensional system model to achieve tractable more efficient computation. Commonly used model order reduction methods such as projection-based proper orthogonal decomposition (POD) \cite{yin2018state}, Hankel-norm optimal solution for multivariable system reduction \cite{kung1981optimal,antoulas2005approximation}, and balanced truncation model methods \cite{gugercin2004survey}. However, these methods typically fall short of preserving the physical meanings of the system's states. In particular, when any of these proposed methods is implemented in an agricultural system, the state of a reduced-order model will no longer accurately reflect the field's soil moisture. Researchers have introduced various techniques for maintaining the topology of the system state in cluster-based model reduction, as discussed in \cite{cheng2019gramian}. Recently, in \cite{sahoo2022knowledge}, a reduced model based on system state trajectories was proposed for an agro-hydrological system. This method can effectively represent the dynamic behavior of a system throughout the entire system operation period. Nonetheless, these approaches are typically implemented for linear systems, and a single reduced-order model is built at the beginning of the season and is used to account for the entire growing season. In \cite{sahoo2022adaptive}, another approach that employs dynamic model reduction and an MHE-based estimator was proposed. An offline heuristic approach based on irrigation prescription is used to change of reduced model over time. This could lead to a substantial discrepancy between the system model and the actual dynamics of soil water, presenting challenges in adapting to real-time conditions that vary over time. Frequent model reduction can also be a consequence of this approach. Additionally, MHE, an online optimization method, generally leads to much higher computational complexity than explicit recursive methods such as EKF. 

In view of the discussed observations, we present a novel method for reducing the complexity of soil moisture state estimation for large-scale agro-hydrological systems, using a performance-triggered adaptive model identification. Our approach involves constructing reduced models that are tailored to the prediction performance of the reduced model to reduce the discrepancy between the ground-truth and the model predictions. The discrepancy can arise from significant disturbances, different soil moisture conditions, and soil parameter variance, among other factors. The proposed approach utilizes the original high-dimensional model as a simulator to generate ground-truth trajectories and an unsupervised machine learning technique is applied to group states that have similar trajectories. The model prediction error over a specified time window is computed to evaluate the prediction discrepancy between the reduced model and the original model. When it exceeds a pre-determined threshold, model re-identification is triggered. An adaptive reduced-order EKF is designed to estimate the soil moisture content in each discrete of the entire field.

The subsequent chapters are organized as follows: Section 2 presents the first-principles model of the agricultural system. Section 3 introduces the proposed adaptive model reduction and state estimation approach. Section 4 showcases the results obtained by applying the proposed adaptive extended Kalman Filter (EKF) to a large-scale agricultural field in various scenarios. Finally, the last section concludes this work.

\section{Preliminaries}
\subsection{Description of agricultural systems}
\begin{figure}
\centering
\includegraphics[width=0.8\textwidth]{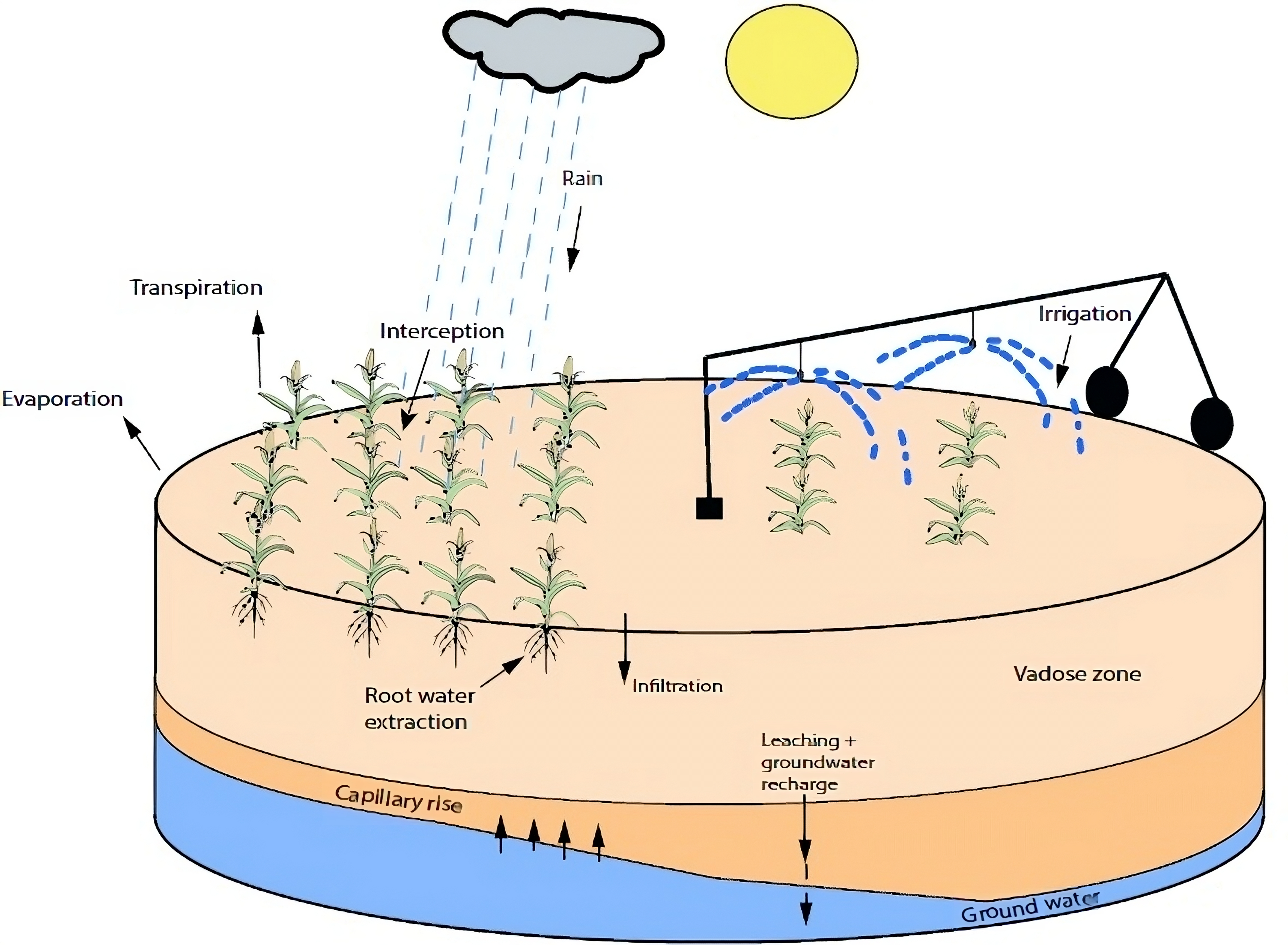}
\caption{A schematic of an agricultural field \cite{agyeman2021soil}}
\label{fig:agro}
\end{figure}
An agro-hydrological system represents the complex interplay between soil, crops, the atmosphere, and water. An illustration of the agro-hydrological systems studied in this work is shown in Figure~\ref{fig:agro}. The water inputs into the system encompass external irrigation to the field and precipitation, while the main water outputs include natural evaporation, transpiration, and groundwater drainage. The process of water infiltration into the soil is driven by both capillary and gravitational forces. We consider agricultural fields equipped with a center-pivot irrigation system, as depicted in Figure \ref{fig:agro}. The center-pivot rotates in a circular pattern. The water dynamics of the field can be modeled using the Richards equation in the cylindrical coordinates as follows \cite{agyeman2021soil}:
\begin{equation}
    c(h)\dfrac{\partial h}{\partial t}= \frac{1}{r}\frac{\partial}{\partial r}\Bigg[rK(h)\frac{\partial h}{\partial r}\Bigg]+
    \frac{1}{r}\frac{\partial}{\partial \theta}\Bigg[\frac{K(h)}{r}\frac{\partial h}{\partial \theta}\Bigg]+
    \frac{\partial}{\partial z}\Bigg[K(h)\Bigg(\frac{\partial h}{\partial z}+1\Bigg)\Bigg]+S(h,z)
\label{eq:cyl_richards}
\end{equation}
where \( h \) [m] represents the pressure head, indicating the height of the water column and $r$, $\theta$, and $z$ represent the spatial variables for radial, azimuthal, and axial directions, respectively. In (\ref{eq:cyl_richards}), \( c(h) \) [m\(^{-1}\)] is the soil water capillary capacity; \( K(h) \) [ms\(^{-1}\)] refers to the unsaturated hydraulic conductivity of the soil, indicating water movement through the soil; \( S(h,z) \) [m\(^3\)m\(^{-3}\)s\(^{-1}\)] represents the sink term, which accounts for water reduction or extraction from the system. A detailed description of the system model can be found \cite{agyeman2021soil}. The Neumann boundary condition characterizes the surface boundary as follows:
\[
\frac{\partial{h}}{\partial{z}}\Bigg|_{r,\theta,z= z_s} = -1 - \frac{u(t)}{K(h)}
\]
where $u(t)$ is the input that includes irrigation and precipitation at the surface of the field, and $z_s$ is the soil depth. The bottom boundary condition of the soil is specified as free discharge.

\subsection{Problem formulation and state-space model}
The Richards equation (\ref{eq:cyl_richards}) is a nonlinear PDE that poses challenges for analytical solutions. For numerical analysis, we adopt the explicit centralized finite difference method to discretize equation (\ref{eq:cyl_richards}) to tackle this challenge. The spatial discretization of the model establishes a continuous-time state-space model as follows:
\begin{subequations}
\label{eq:statespace}
\begin{align}
\dot{x}(t)&= f(x(t), u(t)) + w(t) \label{eq:ss_w}\\ 
y(t)&  = Cx(t) +v(t) \label{eq:ss_v}
\end{align}
\end{subequations}
\noindent where $x(t)\in \mathbb{R}^{N_x}$ denotes the soil pressure head value (state vector) of size $N_x$ and $u \in \mathbb{R}^{N_u}$ represents the irrigation at the surface (input vector) with dimension $N_u$. $y(t)\in \mathbb{R}^{N_y}$ denotes the measurements at each sensor node in pressure head (observation vector), $w(t)\in \mathbb{R}^{N_x}$ is the additive process disturbance, $C$ is a matrix indicating the relation between $x$ and $y$, and $v(t)\in \mathbb{R}^{N_y}$ denotes the noise associated with measurements. 

\begin{figure}
\centering
\includegraphics[width=0.7\textwidth]{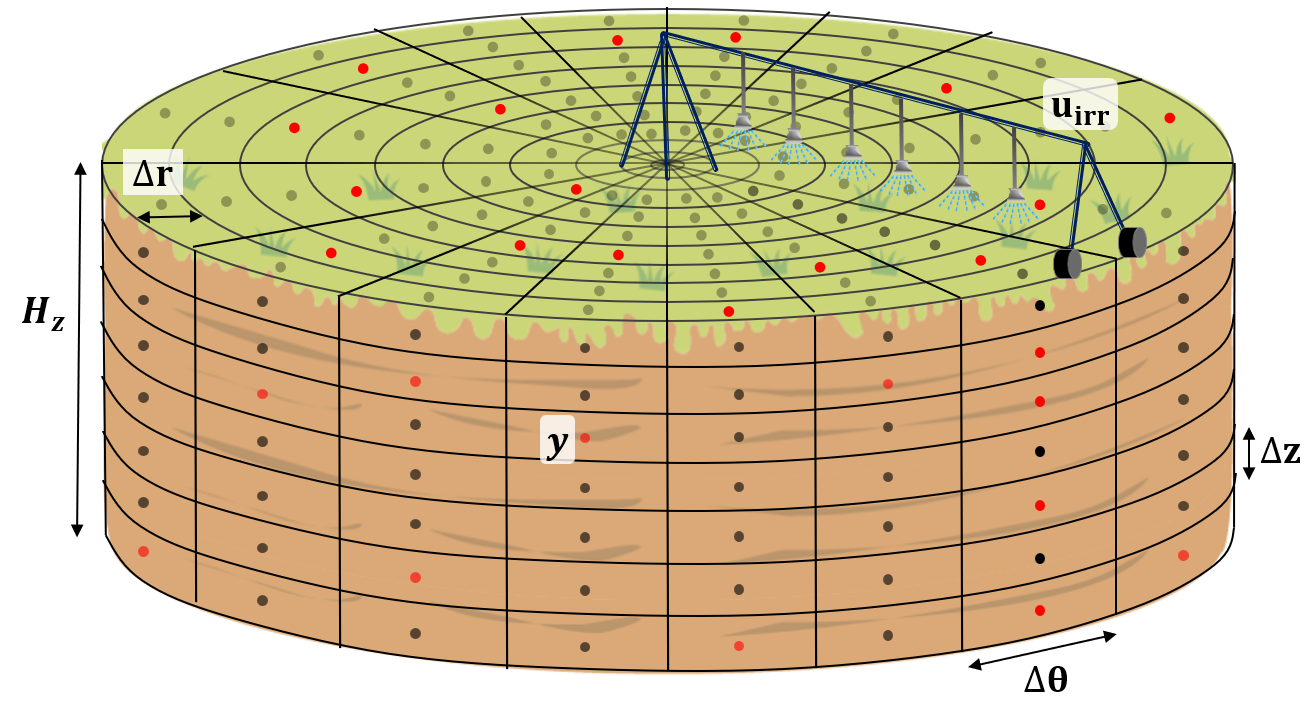}
\caption{Discretization of the agricultural field where each dot denotes the discretized node and red dots indicate the point sensors}
\label{fig:meshagro}
\end{figure}

A discretized diagram of the agricultural field is provided in Figure \ref{fig:meshagro}. The model is discretized into total $N_x$ nodes with $N_r$, $N_{\theta}$, and $N_z$ nodes in the radial, axial, and azimuthal directions, respectively. Noted that the total number of soil moisture (states) is the discretized nodes $N_x$. Therefore, the dimension of the irrigation input $u$ is the same as the radial $N_r$ nodes shown in the diagram (\ref{fig:meshagro}). That means, at any moment, the central pivot can irrigate radial nodes $N_r$ for a particular axial direction, leaving the rest of the field unirrigated. 

\begin{figure}
\centerline{\includegraphics[width=0.3\columnwidth]{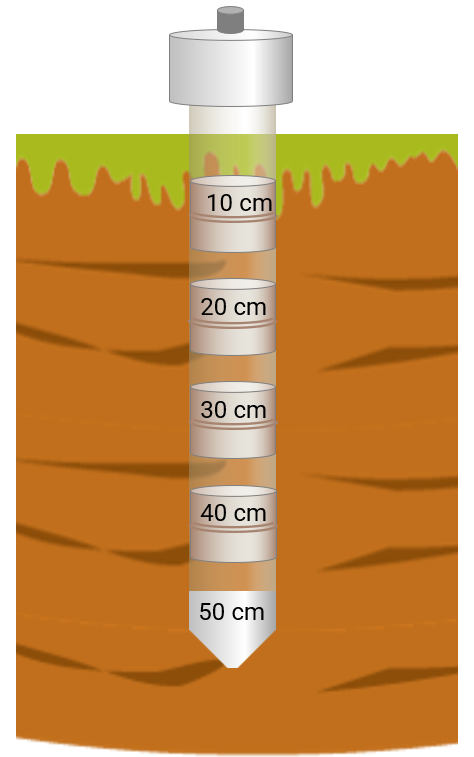}}
\caption{A schematic of a point sensor} \label{fig:point_sensor}
\vspace{-2mm}
\end{figure}

Point sensors are deployed over the field as shown in Figure \ref{fig:point_sensor}. The principle and details of the point soil moisture measurements are provided in \cite{rasheed2022soil}. These sensors measure the soil pressure head at specific locations and depths within the field, such as the surface, the rooting depth, or at various depths within the soil. At each sampling instant, the sensors record the soil pressure head measurements $y$ at a total of $N_y$ locations and the equation (\ref{eq:ss_v}) shows the relation between $x$ and $y$. From a practical point of view, we consider that $N_y \ll N_x$, yet it does not deteriorate the observability of the complete set of states $x$.

Since this system is inherently high-dimensional, the state-space model as shown in (\ref{eq:statespace}), it can be prohibitive to use this model directly as the basis of EKF-based state estimation implementation. To address this issue, we choose to reduce the order of the model to approximate the dynamics of the original system, and use the reduced-order model as the model basis for state estimation design. A reduced-order model that approximates the original model can be obtained and used in state estimation. Over a growing season, soil properties may change over time especially when the soil is excessively dry or wet. Therefore, it is crucial to adapt the reduced model to different conditions. Our goal is to obtain real-time soil moisture information in the form of pressure head $x$ at each discrete node of the field by utilizing the accessible sensor measurements $y$. It is assumed that $y$ is sampled at a sampling time $\Delta$; that is, $y(t_k)$ with $t_k = k\Delta$, $k=0, 1, \ldots$, are available. This is a standard state estimation problem except that the size of $x$ can be very large. The large dimensionality typically leads to two challenges: (a) computational complexity of the model and the associated estimation scheme, and (b) a low degree of observability of soil moisture pressure head ($x$) when the number of measured outputs ($y$) is small. To address this challenge, a performance-triggered adaptive model reduction is designed by using a clustering technique inspired by the work \cite{alanqar2017error}. Finally, we propose a state estimator for soil moisture based on this adaptive model reduction approach.

\section{Proposed model reduction and state estimation} \label{section3}
\begin{figure}
\centerline{\includegraphics[width=0.8\columnwidth]{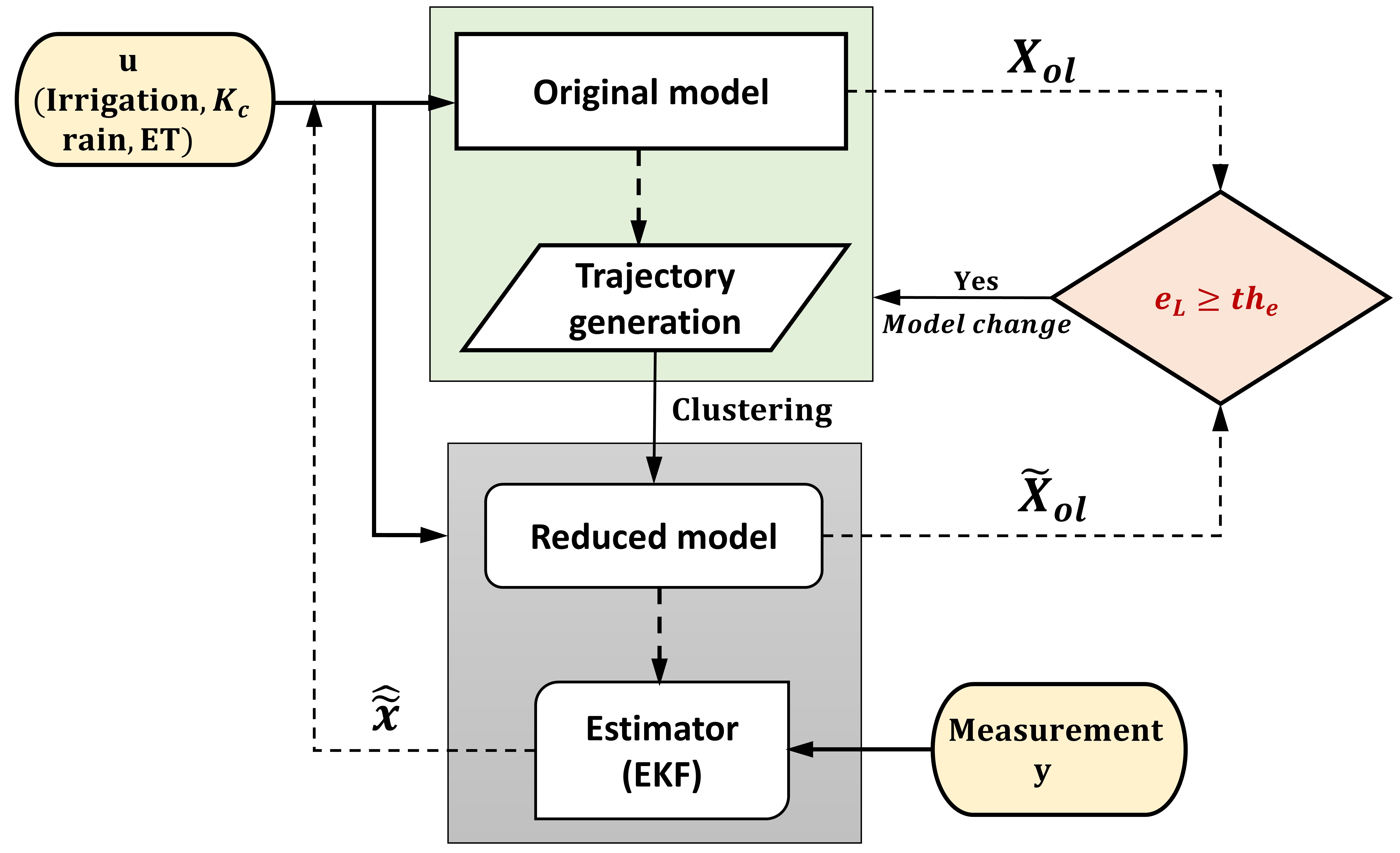}}
\caption{Proposed performance-triggered adaptive model reduction and state estimation scheme} \label{fig:steps}
\end{figure}

In this section, we present the performance-triggered adaptive model reduction method and the associated EKF design. Figure \ref{fig:steps} illustrates the essential steps involved in the proposed approach. The discretized state-space representation of the Richards equation (\ref{eq:statespace}) serves as the original model, and an error metric $e_L$ is evaluated at each sampling time $t_k$ to observe the prediction accuracy of the reduced model against the original model. In Figure \ref{fig:steps}, the algorithm is outlined in four key components: (a) original model for state trajectory generation, (b) reduced-order model for approximating the system dynamics, (c) a criterion for adaptive model reduction $e_L\geq th_e$, and (d) EKF design based on the adaptive reduced-order model. The error metric $e_L$ will be introduced later in equation (\ref{eq:error_metric}) which assesses the performance of the reduced model. In the figure, $X_{ol}$ and $\Tilde{X}_{ol}$ are the open-loop predictions from the original model and reduced model respectively and $\hat {\Tilde x}$ denotes the estimated state using model reduction. If the current reduced-order model begins to fail in describing the soil water dynamics, for instance, due to variations in the soil moisture dynamics, $e_L$ starts to increase, indicating a deviation from the original model. When $e_L$ exceeds a pre-determined threshold $th_e$, the scheduled irrigation input and the weather forecast are used to collect the state trajectories based on simulating the original model, and a new reduced model is created. Soil moisture estimation of the entire field is performed based on the reduced model and field measurements. The main components of the proposed approach are explained in the remainder of this section.

\subsection{Adaptive model reduction}

At each sampling time, the metric $e_L$ is computed, and when $e_L$ it surpasses the threshold $th_e$, the reduced model is re-identified. For model identification initially or upon $e_L$ surpassing the threshold, identical steps are followed, outlined as follows:

\subsubsection*{Step 1: State trajectory generation} 
At a sampling instant $t_k$ when the identification of the reduced-order model is triggered, the state trajectory over a time window is generated. Specifically, the currently estimated state at $t_k$ is used as the initial condition, and the equation~(\ref{eq:statespace}) is simulated with prescribed irrigation actions for total sampling intervals of $N_{f_d}$. The trajectory of the state over the $N_{f_d}$ steps is denoted as $\mathscr{X}_m$ as follows: 
\begin{align*}
    \mathscr{X}_m &=[x(t_{k}) ~x(t_{k+1})~\dots ~x(t_{k+N_{f_d}})]^T
\end{align*}
where $\mathscr{X}_m$ $\in \mathbb{R}^{N_{fd}\times N_{x}}$ is the state snapshot matrix for the $m^{th}$ model reduction assuming that there were $m-1$ model reductions performed before $t_k$. The reduced models that are generated during this process are expected to perform well for at least $N_{f_d}$ sampling time intervals.

\subsubsection*{Step 2: Clustering and reduced model creation} 
An updated reduced-order model is created using the snapshot matrix $\mathscr{X}_m$. Each column of $\mathscr{X}_m$ comprises the trajectory of a state element or a node $x_i$ where $i$ ($i=1,\ldots, N_x$). The purpose of the clustering is to merge similar trajectories into one cluster. Instead of the state element $x_i$, a cluster will be considered as a state element of the reduced model. In this study, we used an agglomerative hierarchical clustering \cite{steinbach_comparison_2000} technique to find the clusters for the trajectories. Initially, individual state elements are treated as clusters, and the distances between these clusters are computed. Subsequently, clusters with distances smaller than threshold ${th}_c$ are consolidated into a new unified cluster. The threshold of the distance between the clusters plays a crucial role as a tunable factor in determining the performance of the reduced-order model. In assessing accuracy, the state elements' similarity is quantified commonly by the Euclidean distance between trajectories or the state elements. The average distance between two clusters is calculated as follows:
\[
\mathcal{D}(a,b) = \frac{1}{n_a n_b}\sum_{m = 1}^{n_a}\sum_{n = 1}^{n_b}d(x_{am},x_{bn}) 
\]
where $a$ and $b$ denote the two distinct clusters, $n_a$, $n_b$ are the sizes of the clusters of $a$ and $b$ respectively, $x_{am}$ and $x_{bn}$ denote data points within clusters $a$ and $b$ respectively. 

Consider that after clustering, there are $r_m$ clusters. Denote by ${C}^{(m)} = \{{C}_1^{(m)},{C}_2^{(m)}, \dots, {C}_{r_m}^{(m)}\}$ the collection of clusters for the $m^{th}$ model reduction. The clusters adhere to the criteria: i) ${C}_i^{(m)} \cap {C}_j^{(m)} = \Phi$ and ii) ${C}_1^{(m)} \cup {C}_2^{(m)} \cup\ldots \cup {C}_{r_m}^{(m)} =\mathscr{X}_m$.

The creation of the \(m^{th}\) reduced system relies on the utilization of the Petrov-Galerkin projection methodology, as elaborated in \cite{antoulas2005approximation}. Within this Petrov-Galerkin projection approach, the fundamental component is the projection matrix, denoted as \({U}^{(m)} \in \mathbb{R}^{N_x\times r_m}\). This matrix is systematically crafted based on the structure of the clusters (\({C}^{(m)}\)). The individual elements of \({U}^{(m)}\) are mathematically represented as follows:
\[\begin{aligned}
{U}^{(m)}_{i,j}=\left\{
\begin{array}{@{}ll@{}}
    w_i, & \text{if}\ \textmd{point}\ i \in {C}^{(m)}_j \\
0, & \text{else}
\end{array}\right.
\end{aligned}
\]
where $\mathit{w_i}$ is the weight of each state element $i$ in a cluster $C_j$ during the $m^{th}$ model reduction and can be found in the following equation:
\[
\begin{aligned}
\mathit{w_{i}}= 1/||\alpha_i||, \
\alpha_i= \mathbb{E}_i ^{T}\alpha
\end{aligned}
\]
where $\alpha$ denotes the inclusion of the state element in the cluster and can be defined as $\alpha = [1,\dots,1]^T \in \mathbb{R}^{N_x}$, $||\alpha_i||$ denotes the $L_2$ norm of $\alpha_i$, $\mathbb{E}_i = e_{{C}_i}\in \mathbb{R}^{N_x\times N_{C_i}}$ ($N_{C_i}$ denied as the size of cluster $i$ $C_i$) is a matrix with columns of $e_j$'s and each $e_j$ is the $j$-th column of the identity matrix $\mathbb{R}^{N_x\times N_x}$.

\begin{figure}
\centerline{\includegraphics[width=0.6\columnwidth]{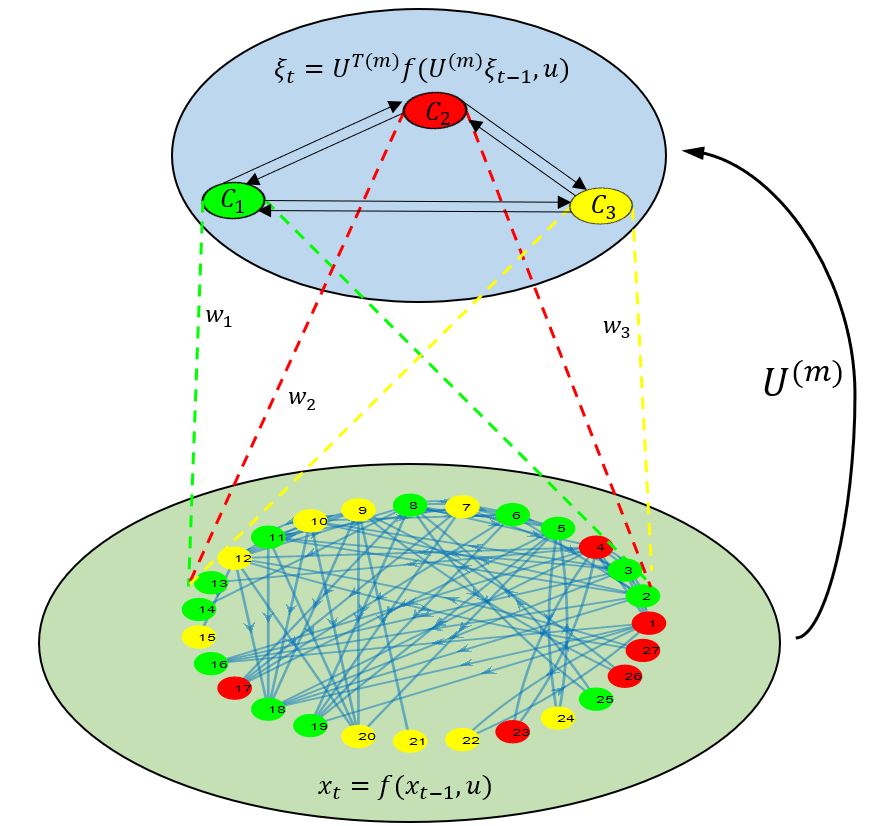}}
\caption{A representation of $m^{th}$ model reduction} \label{fig:reduced}
\vspace{-2mm}
\end{figure}
The $m^{th}$ reduced model of (\ref{eq:statespace}) is illustrated in Figure~\ref{fig:reduced}, and the reduced-order state-space model can be expressed as:
\begin{equation}
\dot{\xi}^{(m)}(t)= f_r^{(m)}(\xi^{(m)}(t), u(t), w(t)) \label{eq:red_nonlinear}
\end{equation}
where $f_{r}^{(m)}= {U^{(m)}}^{\text{T}}f$ and $\xi^{(m)}(t) = {U^{(m)}}^{\text{T}}x(t)$. After carrying out the numerical discretization, we found a discrete-time reduced model is shown below: 
\begin{equation}
\xi^{(m)}(t_{k+1})= f_{rd}^{(m)}(\xi^{(m)}(t_{k}), u(t_{k}), w(t_{k})) \label{eq:red_nonlineard}
\end{equation}
where $f_{rd}$ is the discrete-time function of the reduced-order model. An approximation of the state x can be obtained based on the state prediction of the reduced-order model as $\Tilde{x}(t) = {U}^{(m)}\xi^{(m)}$.
\begin{figure}
\centerline{\includegraphics[width=0.7\columnwidth]{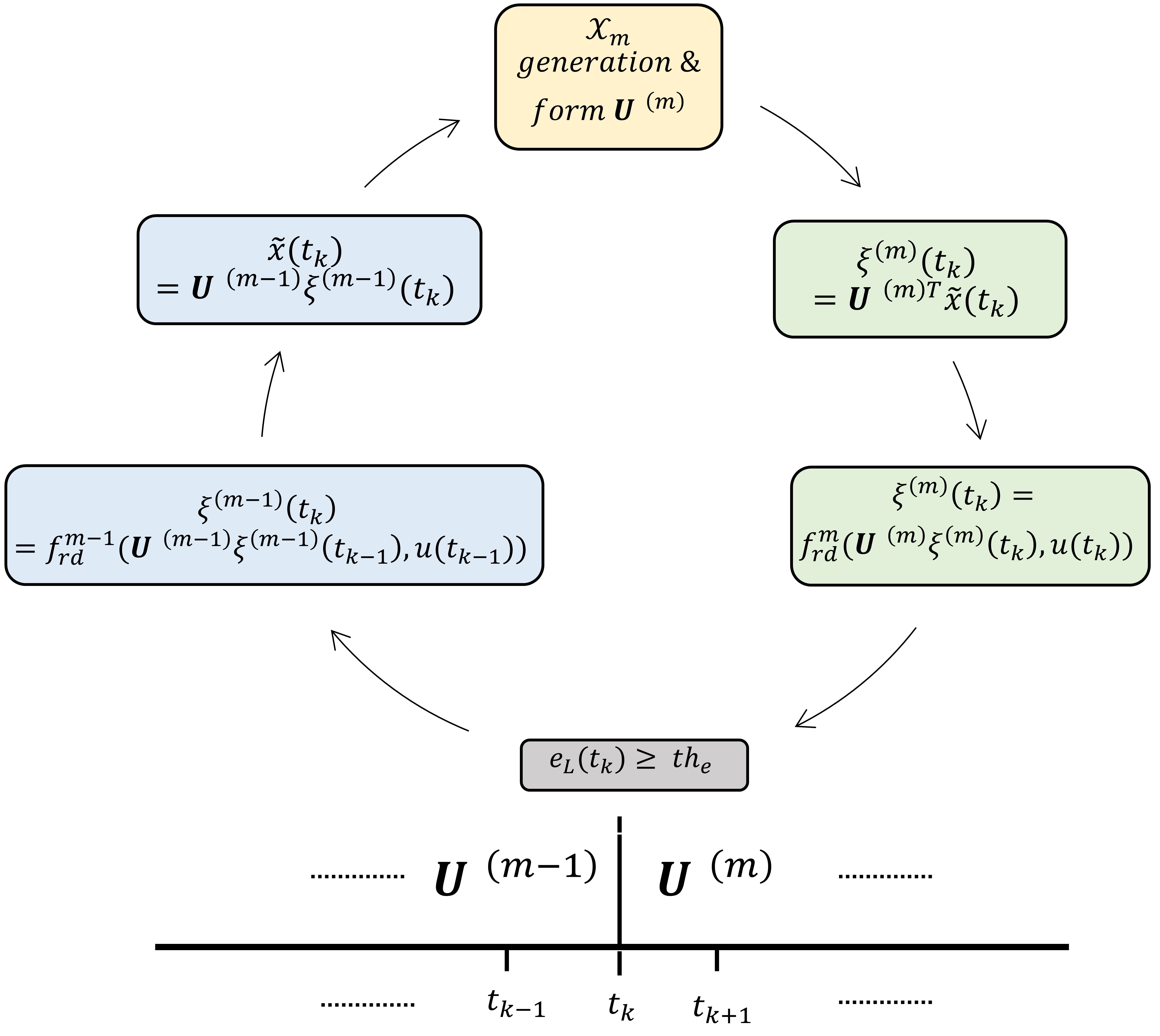}}
\caption{Information transformation from one reduced model to another} \label{fig:step_4}
\end{figure}
It is important to note that during the transition from one reduced model to another, the dimension of the new reduced-order model may change. To facilitate a seamless transition between these reduced models, a two-step process is employed. First, the state information of the previous reduced model is mapped back to the full state space. Second, the entire state space is projected onto the new reduced model using the recently computed projection matrix. Figure \ref{fig:step_4} illustrates how information is transferred between models.

\subsection{Adaptive extended Kalman filter}
We propose an adaptive reduced-order state estimator using EKF. EKF is a commonly utilized method for state estimation of nonlinear systems, characterized by its process of consecutively linearizing the nonlinear system at each step. The advantage of using EKF is its computational efficiency \cite{debnath2022subsystem}. 

As explained earlier, when there is a model update or re-identification, the dimension of the new reduced-order model may not remain the same. Therefore, the standard EKF cannot be applied. To address this issue, we develop an adaptive EKF based on the adaptive reduced-order model. Let us define the discrete reduced-order model (\ref{eq:red_nonlineard}) with the corresponding output equation as follows:
\begin{equation}
    \begin{aligned}
        \xi^{(m)}(t_{k+1}) & =  f_{rd}^{(m)}(\xi^{(m)}(t_{k}),u(t_{k}), w(t_{k})) \\
        y(t_{k}) & =  C_r^{(m)}\xi^{(m)}(t_{k})+v(t_{k})  
\end{aligned}
\label{eq:red_system_w}
\end{equation}
where $v(t_{k})$ denotes the measurement noise at time $t_k$, $C_r^{(m)}=CU^{(m)}$. 

There are two steps in EKF: the prediction and the update of the states. At the sampling time $t_{k}$, in the prediction step, the adaptive EKF first predicts $\xi^{m}$ at $t_k$ based on the state estimate at $t_{k-1}$ and the reduced model as follows:
\begin{equation}
\hat {\xi}^{(m)}(t_{k|k-1})= f_{rd}^{(m)}(\hat \xi^{(m)}(t_{k-1}),u(t_{k-1}), w(t_{k-1}))  \label{eq:EKF_p1}
\end{equation}
where $\hat {\xi}^{(m)}(t_{k|k-1})$ represents the  reduced state prediction at time instant $t_k$ using an initial guess $\hat {\xi}^{(m)}(t_{0})$ or the previously estimated reduced state $\hat {\xi}^{(m)}(t_{k-1})$. The evolution of the variance of the reduced states is also propagated based on the reduced model:
\begin{equation}
P_r^{(m)}(t_{k|k-1})= A_{d}^{(m)}(t_{k-1}) P_r^{(m)}(t_{k-1}) A_d^{(m)}(t_{k-1})+Q_r^{(m)} \label{eq:EKF_p2}
\end{equation}
where $P_r^{(m)}$ and $Q_r^{(m)}$ denote covariance matrices for the state and process disturbances in reduced-order form, respectively, and $A_{d}^{(m)}(t_{k-1})=\frac {\partial f_{rd}^{(m)}}{\partial \xi^{(m)}}\bigg |_{\hat{\xi}^{(m)}(t_{k-1})}$ is the state-transition matrix obtained by linearizing the nonlinear reduced model at the estimated state at $t_{k-1}$. Denote by $P(t_0)$ and $Q$ the covariance matrices of the initial state and process disturbances for the original system, we have that $P_r^{(1)}= {U^{(1)}}^{\text{T}}P(t_0){U}^{(1)}$ and $Q_r^{(m)}= {U^{(m)}}^{\text{T}}Q{U}^{(m)}$. 

At each sampling instant $t_{k}$, the measurement $y(t_k)$ is used to update the predictions generated in the prediction step. In the update step of EKF, the current reduced state estimate $\hat {\xi}^{(m)}(t_{k})$ is calculated based on the predicted value $\hat {\xi}^{(m)}(t_{k|k-1})$ as follows:
\begin{equation}
        \hat {\xi}^{(m)}(t_{k})= \hat {\xi}^{(m)}(t_{k|k-1}) + K_r^{(m)}(t_k)(y(t_{k})- C_r^{(m)} \hat {\xi}^{(m)}(t_{k|k-1}))
\label{eq:EKF_p3}
\end{equation}
where $\hat {\xi}^{(m)}(t_{k})$ is the estimated reduced state at time $t_k$, and $K_r^{(m)}(t_k)$ is the correction gain which minimizes a \textit{posteriori} error covariance using the observation innovation $y(t_k)-C_r^{(m)} \hat {\xi}^{(m)}(t_{k|k-1})$. The correction gain is determined as follows: 
\begin{equation}
    K_r^{(m)}(t_k) = P_r^{(m)}(t_{k|k-1}) C_r^{(m)T} (R +  C_r^{(m)} P_r^{(m)}(t_{k|k-1}) C_r^{(m)T})^{-1}
\label{eq:EKF_p4}
\end{equation}
Here, $R$ represents the covariance matrix for observation noise. Additionally, the process involves updating the covariance matrix of the system state in the following equation:
\begin{equation}
    P_r^{(m)}(t_{k})=(I_{r_m}- K_r^{(m)}(t_k)C_r^{(m)})P_r^{(m)}(t_{k|k-1})
    \label{eq:EKF_p5}
\end{equation}
where $P_r^{(m)}(t_{k})$ denotes the \textit{posteriori} error covariance matrix related to the state estimation error at time $t_k$ and $I_{r_m}$, denoted by an identity matrix specific to the $m^{th}$ reduced model with size $r_m$. It is noted that $P(t_0)$, $Q$, and $R$ are tuning parameters for the EKF. The state estimate at the time $t_k$, denoted by $\hat {\Tilde{x}}$, is evaluated following
\begin{equation}
    \hat {\Tilde{x}} (t_k)= {U}^{(m)} \hat {\xi}^{(m)}(t_k)
    \label{eq:EKF_p6}
\end{equation}

When there is a model update, the information in the EKF estimator related to the previous model should be smoothly transferred to the EKF built on the new reduced model. The information transfer is performed by mapping all the information back to the full state system and then projecting it to the new reduced model. Consider that we need to transfer the information of the EKF based on the $m^{th}$ reduced model to the EKF built on the $(m+1)^{th}$ model. The following steps are performed:
\begin{itemize}
    \item Mapping the estimated reduced state and state covariance to the full order state and covariance: $\hat {\Tilde{x}}= {U}^{(m)} \hat {\xi}^{(m)}$, $P={U}^{(m)} P_r^{(m)}{U^{(m)}}^{\text{T}}$.
    \item Projecting the full order information to the new reduced model using the new projection matrix: $\hat {\xi}^{(m+1)} = {U^{(m+1)}}^{\text{T}}\hat {\Tilde{x}}$, $P_r^{(m+1)}= {{U}^{(m+1)}}^{\text{T}}P{U}^{(m+1)}$.  
\end{itemize}

\subsection{Design of the error metric $e_L$ and implementation algorithm}

The reduced model update or re-identification is triggered by an error metric $e_L$, which is evaluated every sampling time to assess the performance of the reduced model. Specifically, at $t_k$, the estimated state $\hat {\Tilde{x}}(t_k)$ is considered as the initial condition. The trajectory of the system state over the next $N_{f_d}$ steps is predicted both based on the original model of (\ref{eq:statespace}) and the current reduced model. An open-loop irrigation profile for a window of $N_{f_d}$ is considered to evaluate the performance of the reduced model. After generating the predictions, the deviation between the trajectory generated by the reduced model and the trajectory of the original model is calculated and used as the error metric $e_L$. The design of $e_L$ is inspired by the work of \cite{alanqar2017error} and is shown below: 
\begin{align}
    e_L(t_k) = \dfrac{1}{N_x}\sum\limits_{j=1}^{N_{f_d}}\sum\limits_{i=1}^{N_x} |\tilde x_i(t_{k+j})- x_i(t_{k+j})|
   \label{eq:error_metric}
\end{align}
where $\tilde x_{i}$ and $x_{i}$ denote the predictions of $i^{th}$ state element obtained based on the reduced model and the original model, respectively. A prediction horizon $N_{f_d}$ is considered. A model reduction is triggered if $e_L(t_k)$ exceeds the predefined threshold $th_e$. A new reduced model is generated as discussed earlier. 
In Algorithm \ref{Adapt_alg}, it employs the metric $e_L$ and $\dot{e}_L$ as the criteria for model adjustments. The current value of $\dot{e}_L(k)$ is derived from the moving average calculated over the preceding ten consecutive differences of $e_L$ when $e_L$ is on the rise. This is used to filter out model disturbances that could affect model change decisions. To generate a snapshot matrix for cluster formation, the algorithm applies an initial soil content, $u_{ir}$, rain, evapotranspiration ($ET$), and crop coefficient ($K_c$) to the system (\ref{eq:statespace}). The evapotranspiration rate ($ET$) characterizes the water loss from a cultivated surface, while the crop evapotranspiration coefficient ($K_c$) is employed to quantify the crop's water use. For further details, one can refer to \cite{agyeman2021soil}. The estimation process takes place in a reduced space, with the necessary variables transformed using the projection matrix. By utilizing the EKF design, from (\ref{eq:EKF_p1}) to (\ref{eq:EKF_p6}), the algorithm computes estimates for soil moisture. By incorporating these strategies, the algorithm achieves efficient computation, reduces model updates when it is needed based on model assessment, and provides accurate estimations of soil moisture.

The proposed reduced-order state estimation is summarized in the following Algorithm \ref{Adapt_alg}:
\begin{algorithm}[ht!]
\caption{Performance-triggered reduced EKF algorithm}
\label{Adapt_alg}
\begin{algorithmic}[1]

\Require Initial guess $\hat{x}_{0}$, $P(0)>0$, $Q, R>0$, $th_e$ ,$th_C$, $m=0$, and $N_{f_d}$
\For{$k = 0 \dots n$}
\If{\big($e_L(k) > th_{e}$ $ \wedge $ $\dot {e}_L(k) \geq 0.05  \vee k==0$\big)}
\State $m \leftarrow m+1$
\State  Apply $\hat{\Tilde{x}}_{k|k}$ and input to generate $\mathscr X_m$ and obtain $U^{(m)}$
\State  Convert $\hat{\xi}^{k|k} = U^{(m)}\hat{\Tilde{x}}_{k|k}$, $P_r^{(m)}(k) = {U^{(m)}}^{\text{T}}P(k) U^{(m)}$ and $Q_r^{(m)} = {U^{(m)}}^{\text{T}}Q U^{(m)}$

\EndIf
\State  Obtain measurements $y(k)$
\State  Calculate current reduced estimates $\hat{\xi}^{(m)}(k|k)$ 
\State  Convert to actual state $\hat{\Tilde{x}}_{k|k}= U^{(m)}\hat{\xi}^{(m)}(k|k)$
\State  Compute $e_L(k)$ and $\dot {e}_L(k)$ 
\EndFor
\end{algorithmic}
\end{algorithm}

\begin{figure}
\centering
\includegraphics[width=1.0\columnwidth]{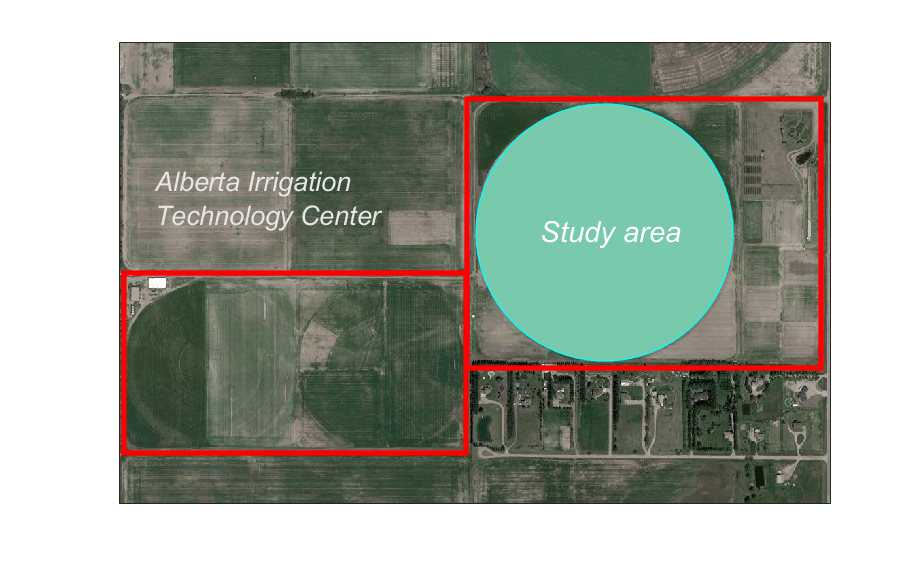}
\caption{Investigated area in Lethbridge, Alberta, Canada
\cite{agyeman2022simultaneous}}
\label{fig:demofarm}
\end{figure}
\section{Application to an agricultural field}

In this section, we present a demonstration of the efficacy of the proposed adaptive model reduction and state estimation in the soil moisture estimation of a large-scale agricultural farm. 

\subsection{Simulation settings}
\begin{figure}
\centering
\includegraphics[width=1.0\columnwidth]{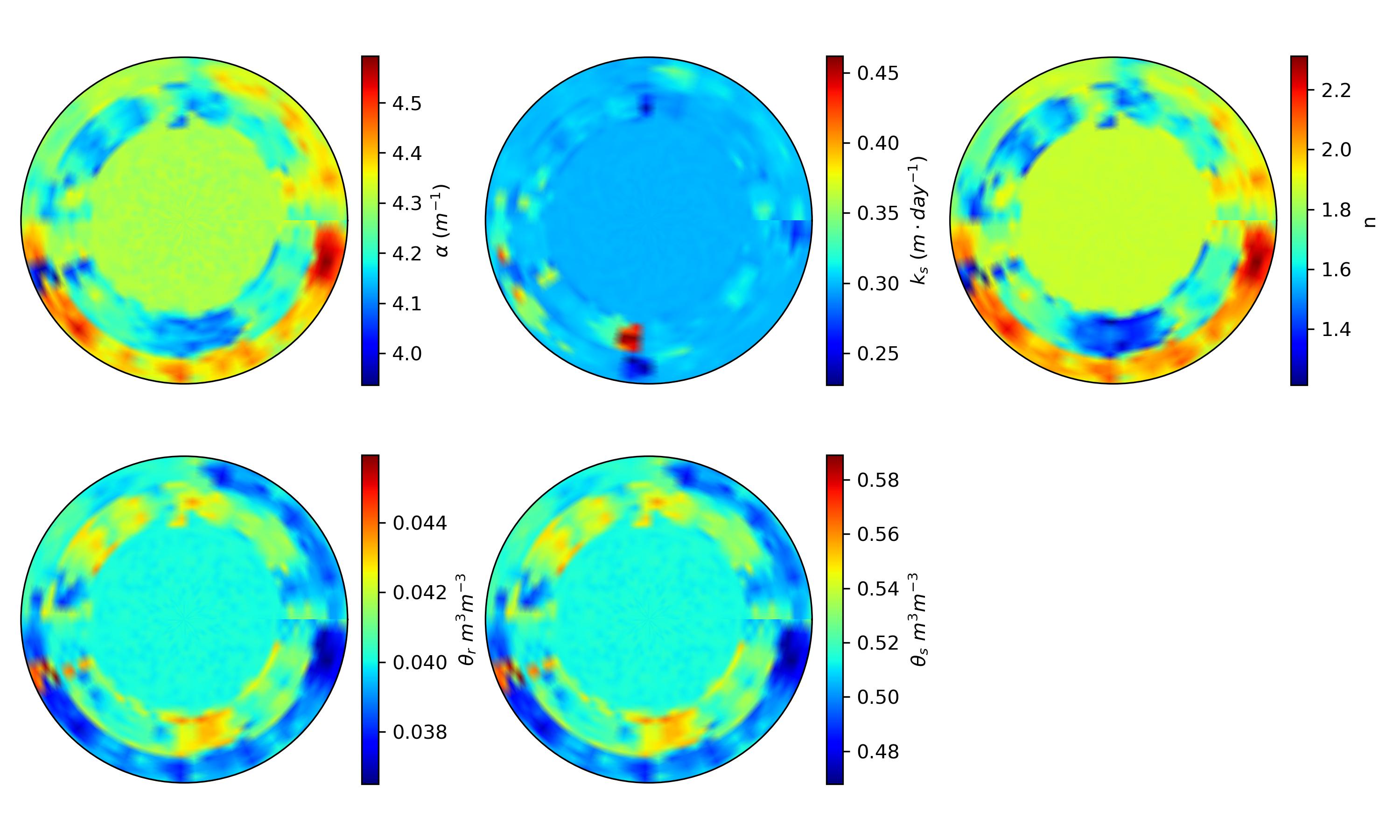}
\caption{Different soil parameters ( $\alpha$, $K_s$, $n$, $\theta_r$, and $\theta_s$) used for the study
\cite{sahoo2022adaptive}}
\label{fig:parameter1}
\end{figure}

The investigated field is a circular field measuring $26.4$ hectares, situated at the Alberta Irrigation Technology Center in Lethbridge, southern Alberta at latitude $49.72$ N and longitude $112.80$ W. The research farm, highlighted in green in Figure \ref{fig:demofarm}, serves as the study area. Within this expansive field, a five-span center-pivot irrigation system is deployed. Furthermore, the field is facilitated with a commercially viable irrigation system with variable rates. The soil hydraulic parameters used in this study are provided in Figure \ref{fig:parameter1}.

\begin{figure}
\centering
\includegraphics[width=0.6\columnwidth]{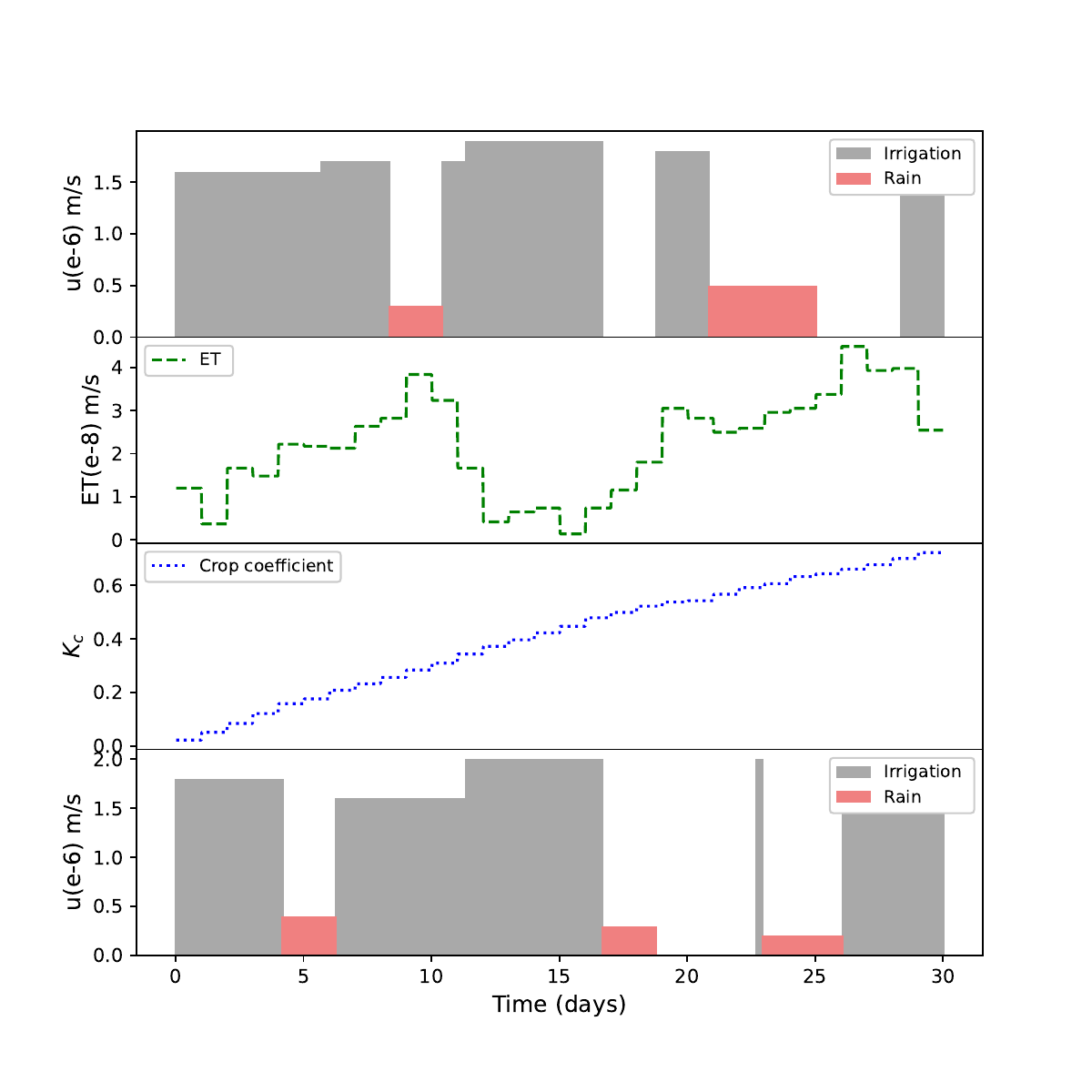}
\caption{Input profile (irrigation, ET, $K_c$, rain) of the system: real-time irrigation and rain information (top) and scheduling and forecast error irrigation and rain disturbance (bottom)}
\label{fig:input}
\end{figure}

 In this study, the soil depth is set as 0.4 m and is evenly divided into 12 discrete nodes. The field has a radius of $290$ m, which is discretized radially into $30$ nodes and azimuthally into $68$ nodes. In this work, we assume the use of point sensors to provide measurements of the soil pressure head values at $90$ selected nodes of the farm at each sampling time, including surface nodes and nodes at various depths. Readers are encouraged to explore the optimal placement of soil sensors for accurate state estimation, as discussed in the work \cite{sahoo_optimal_2019}. The total number of discretized nodes (states) in the research field is $20400$. To simulate the system, information about $ET$, $K_c$, irrigation, and rain is used, as shown in Figure \ref{fig:input}. The initial actual soil moisture $x_0$ in pressure head ($\text{m}$) is simulated to be a distinct value for each quadrant of the field, namely, $-13.5$, $-14.0$, $-12.7$, and $-11.5$ $\text{m}$.

\begin{figure}
\centering
\includegraphics[width=0.75\columnwidth]{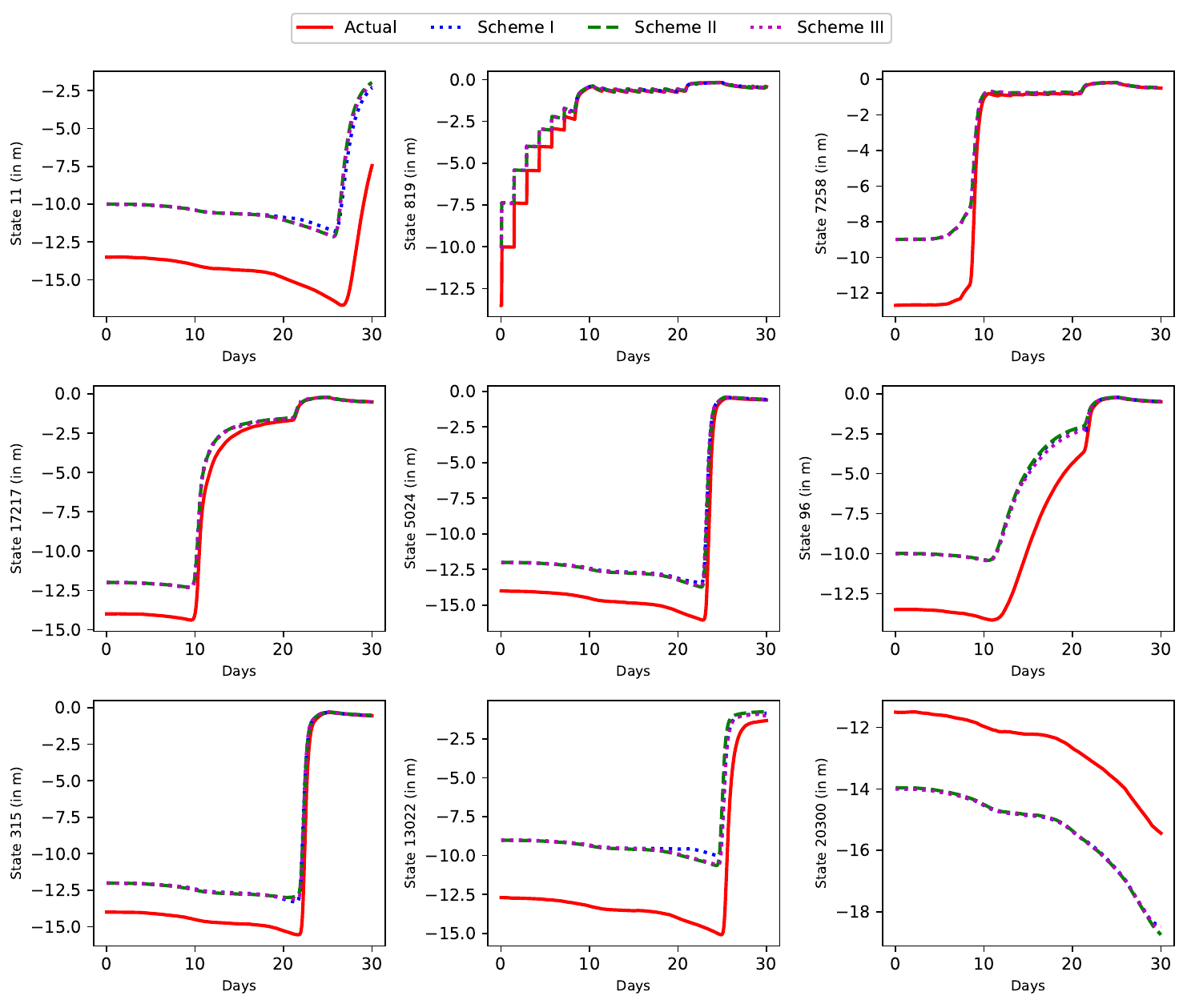}
\caption{Actual state trajectories and state prediction of schemes I, II, and III for large field}
\label{fig:model_perform_a}
\end{figure}
\begin{figure}
\centering
\includegraphics[width=0.75\columnwidth]{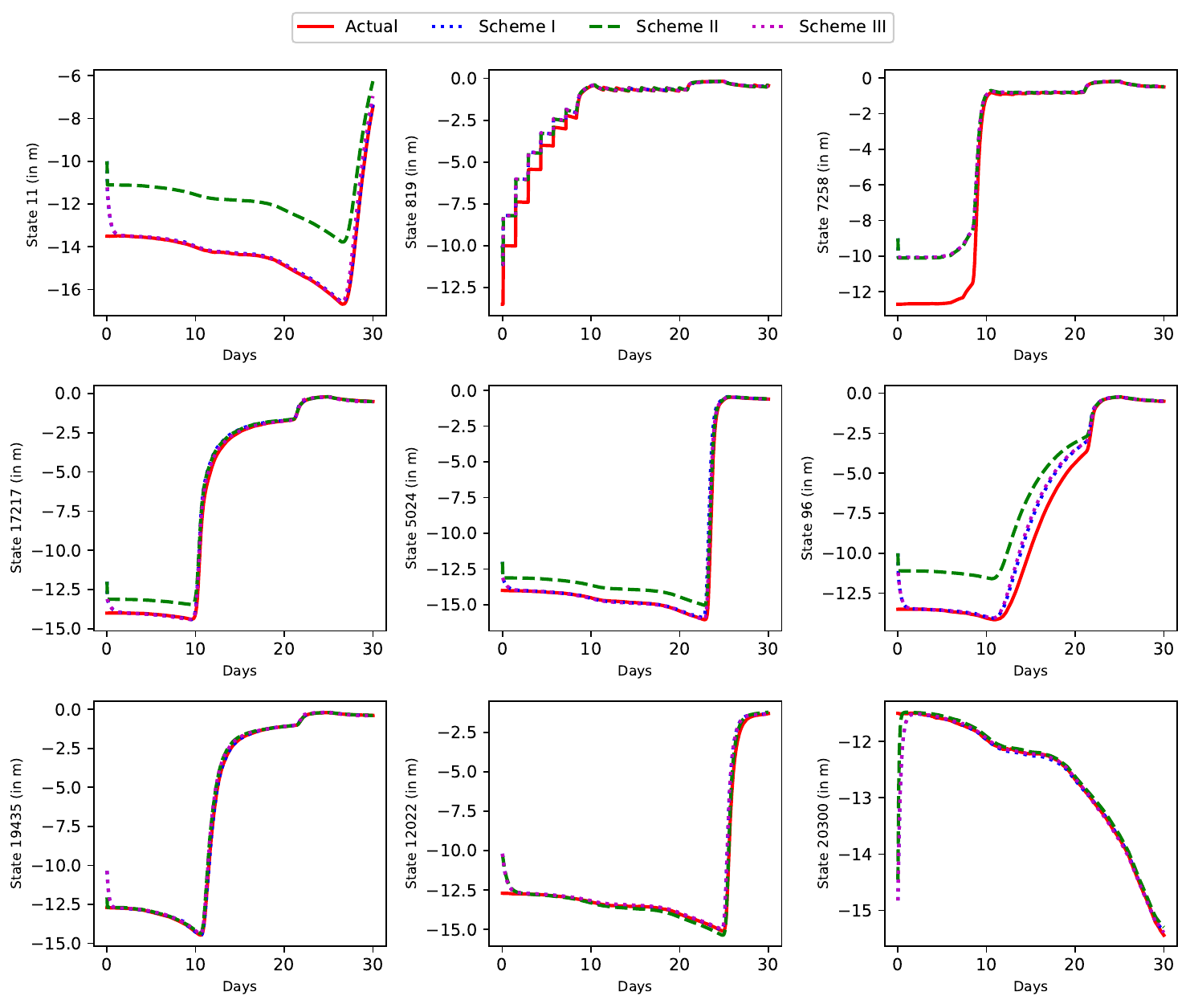}
\caption{Actual state trajectories and the state estimation of schemes I, II, and III for large field}
\label{fig:EKF_perform_a}
\end{figure}
\begin{figure}
    \centering
    \begin{subfigure}[b]{0.5\linewidth}
        \centering
        \includegraphics[width=1.00\columnwidth]{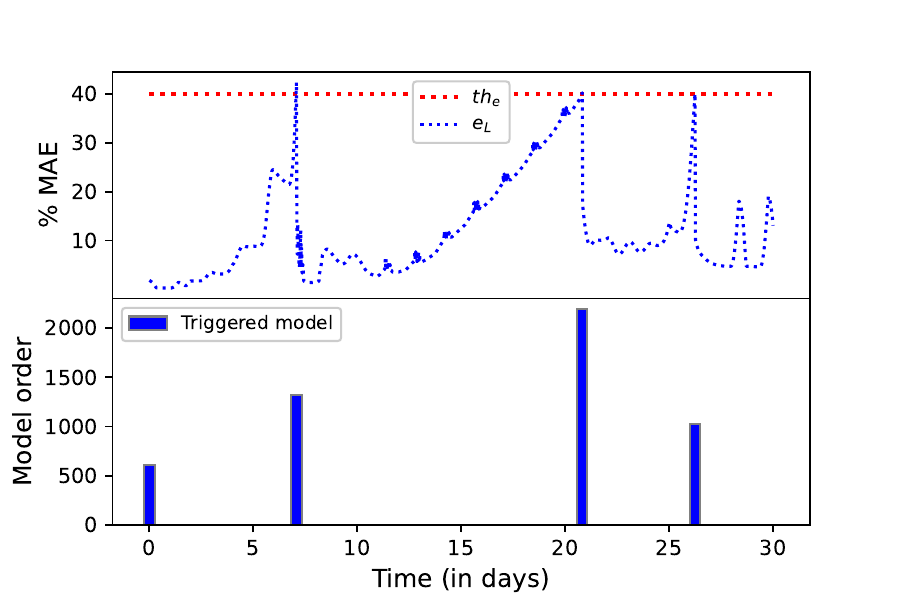}
        \caption{$N_{f_d}=250$}
        \label{fig:scheme1_a}
    \end{subfigure}\hfill
    \begin{subfigure}[b]{0.5\linewidth}
        \centering
        \includegraphics[width=1.00\columnwidth]{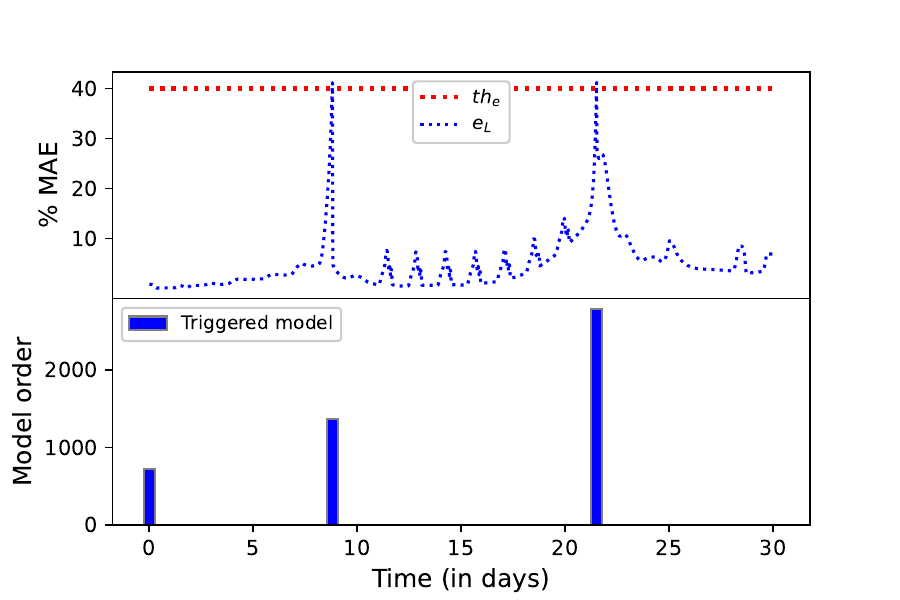}
        \caption{$N_{f_d}=350$}
        \label{fig:scheme1_a1}
    \end{subfigure}
    \caption{The proposed performance-triggered adaptive EKF Scheme I: change of \% MAE  for the reduced model (top) and model re-identification instances with model orders (bottom)}
    \label{fig:scheme1}
\end{figure}

\begin{figure}
    \centering
    \begin{subfigure}{1.0\linewidth}
        \centering
        \includegraphics[width=0.85\columnwidth]{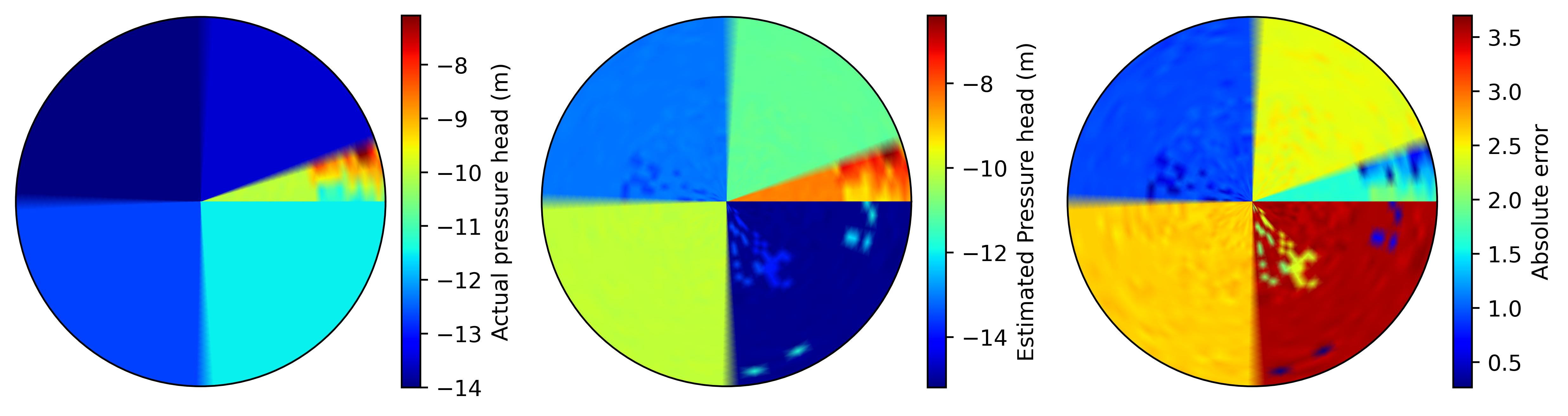}
        \caption{Initial surface soil water map on day one}
    \end{subfigure}%
    \\
    \begin{subfigure}{1\linewidth}
        \centering
        \includegraphics[width=0.85\columnwidth]{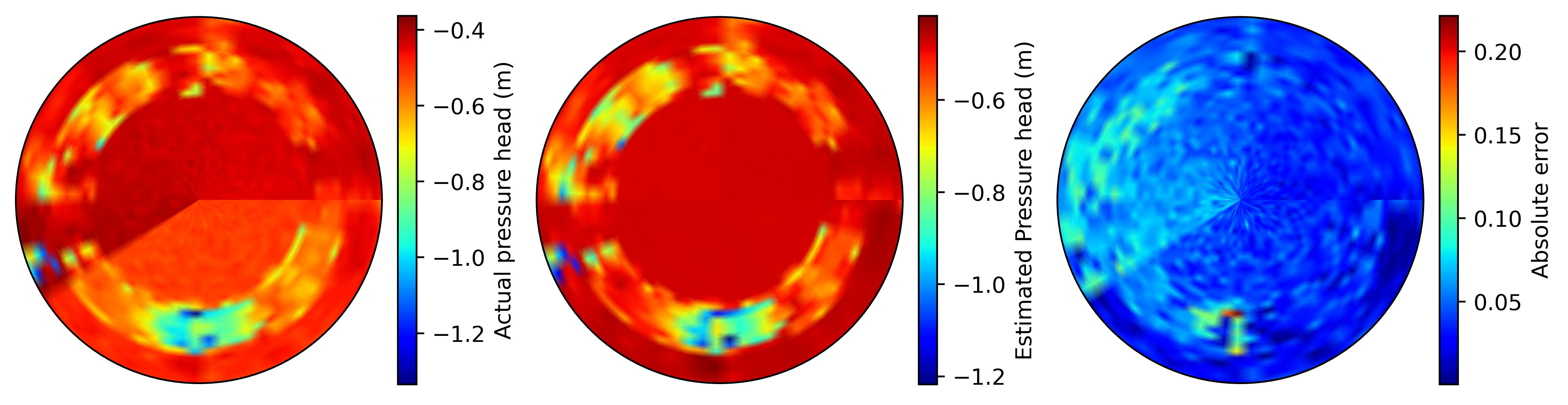}
        \caption{  Final surface soil water map on the last day}
    \end{subfigure}
    \caption{Soil moisture pressure head distribution for actual, estimated, and absolute error between actual and estimated states for surface (Left to right)}
    \label{surface_map}
\end{figure}
\begin{figure}
    \centering
    \begin{subfigure}{1.0\linewidth}
        \centering
        \includegraphics[width=0.85\columnwidth]{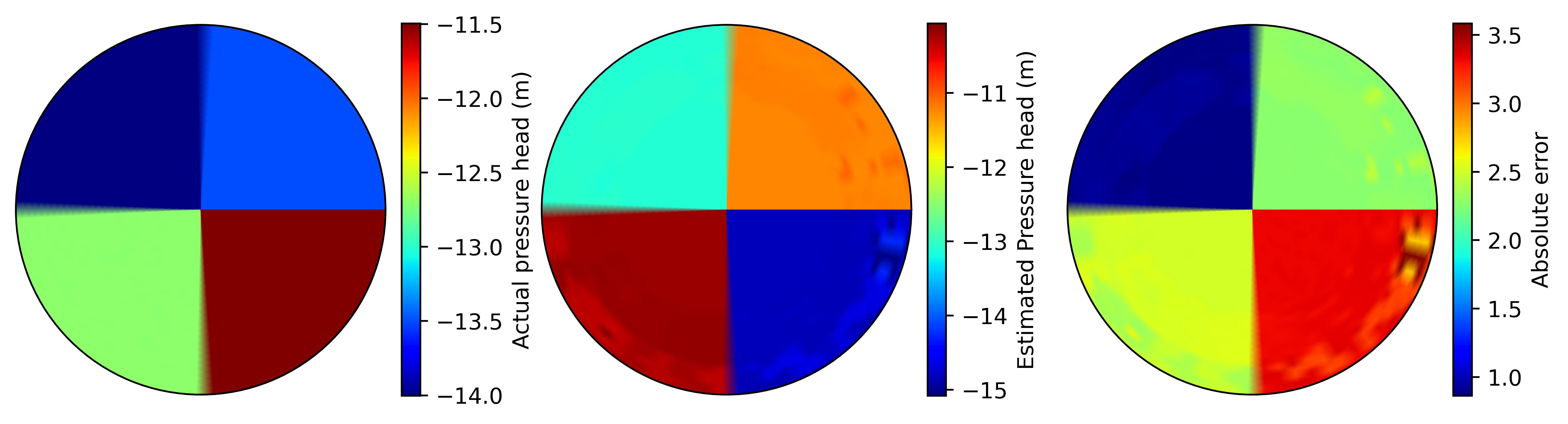}
        \caption{Initial soil water map on day one}
    \end{subfigure}%
    \\
    \begin{subfigure}{1\linewidth}
        \centering
        \includegraphics[width=0.85\columnwidth]{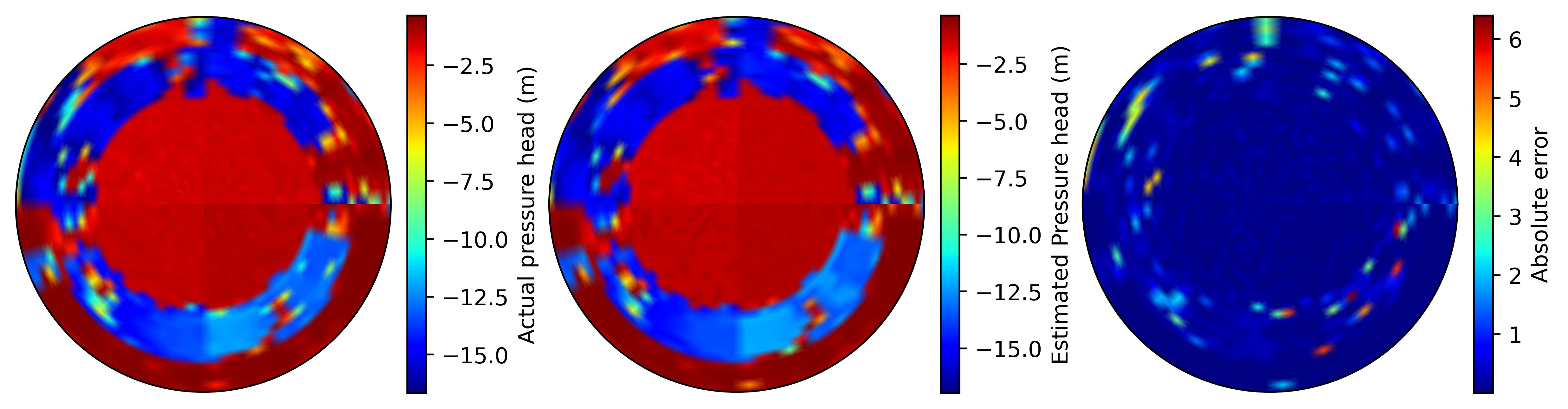}
        \caption{ Final soil water map on the last day}
    \end{subfigure}
    \caption{Soil moisture pressure head distribution at $30$ cm for actual, estimated, and absolute estimation error between actual and estimated states (Left to right)}
    \label{rootzone_map}
\end{figure}

\begin{figure}
    \centering
    \begin{subfigure}[b]{0.5\linewidth}
        \centering
        \includegraphics[width=1.00\columnwidth]{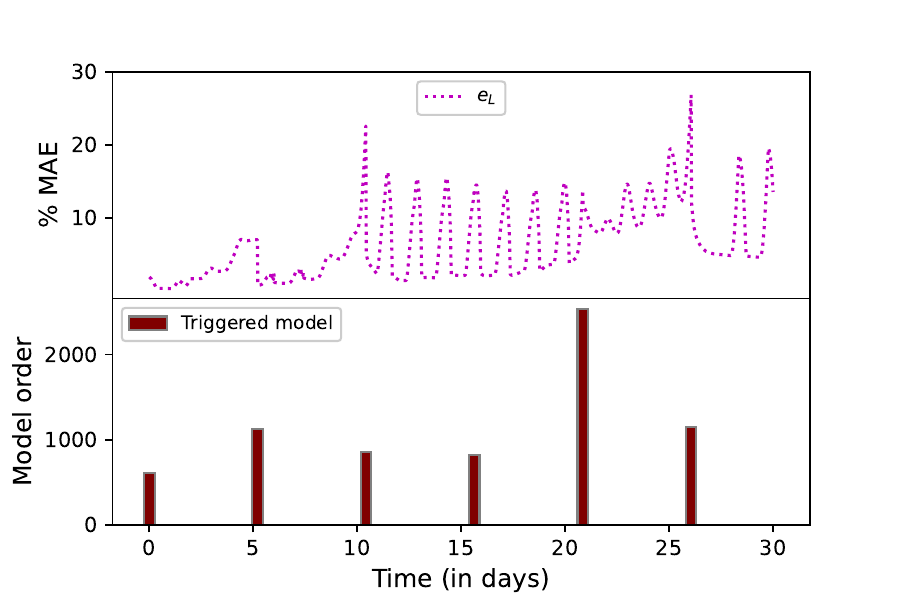}
        \caption{$N_{f_d}=250$}
        \label{fig:scheme3_a}
    \end{subfigure}\hfill%
    \begin{subfigure}[b]{0.5\linewidth}
        \centering
        \includegraphics[width=1.00\columnwidth]{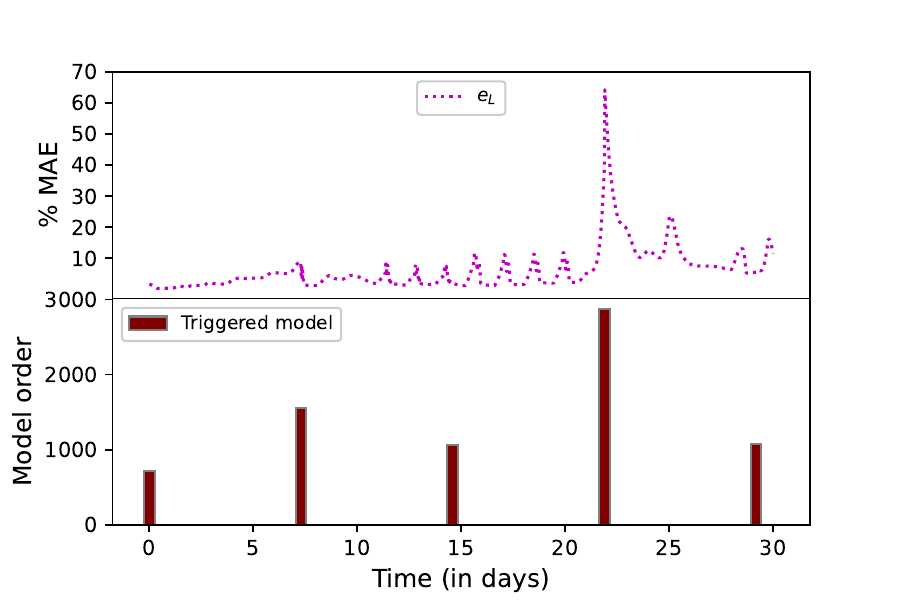}
        \caption{$N_{f_d}=350$}
        \label{fig:scheme3_a1}
    \end{subfigure}
    \caption{The time-triggered adaptive EKF Scheme III: change of \% MAE  for the reduced model (top) and model re-identification instances with model orders (bottom)}
    \label{fig:scheme3}
\end{figure}

The simulation runs for a total of $30$ days with a sampling time $\Delta = 30~\text{min}$, and the initial soil moisture $x(0)$ is unknown and taken as an initial guess $\hat {x}(0)$ as $-10.0$, $-12.0$, $-9.0$, and $-14.0$ $\text{m}$ for the four quadrants, which differ from the actual values of soil moisture $x(0)$. The units for the soil pressure head values are measured in meters ($\text{m}$). The covariance matrices $R$ and $Q$ are defined as diagonal matrices with diagonal elements of $R= 0.08 \times I_{N_y}$ and $Q= 1.0 \times I_{N_x}$, respectively. $I_{N_y}$ and $I_{N_x}$ are the identity matrices of dimension $N_y$ and $N_x$ respectively. The initial state covariance matrix $P$ has entries of $5 \times 10^{-5}$ for all off-diagonal elements and a value of $1.0$ for all diagonal elements. The process disturbances and measurement noise are generated following normal distributions with a zero mean and variances of $10^{-7}$ and $0.8$, respectively.

We explore three estimation schemes to demonstrate the effectiveness of the proposed approach. Scheme I is based on the proposed performance-triggered model reduction and EKF design. In Scheme II, a reduced EKF is designed based on a single non-adaptive reduced model for the total simulation time. To reduce input disturbance in Scheme II, a reduced EKF is designed based on a time-triggered adaptive reduced model in Scheme III. We assume that all the information, including irrigation, $ET$, $K_c$, and rain, is available for the total operation duration accurately. The conventional EKF based on the original nonlinear system model is not included in this study due to its computational infeasibility limitation for the large-scale system, as discussed in \cite{sarupa_2023}.
\begin{figure}
\centering
\includegraphics[width=0.7\columnwidth]{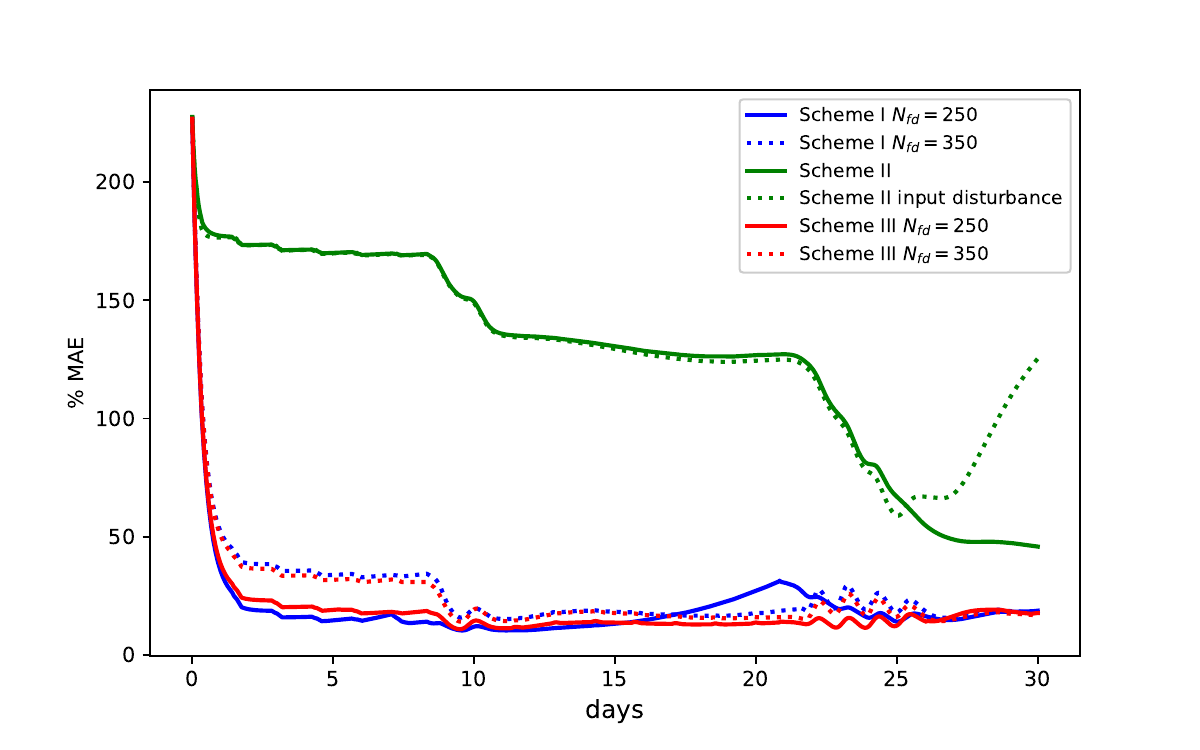}
\caption{State estimation performance (\% MAE) of all the schemes for large field}
\label{fig:scheme2_a}
\end{figure}
Note that the initial reduced model is created by applying the initial guess $\hat x({0})$ as the initial state and not the actual initial state $x(0)$ which is unknown for all the schemes. The purpose of the estimator is to estimate the initial soil moisture of the agricultural system. Considering $x(0)$ as the actual state at $t_0=0$, actual trajectories are generated to compare the reduced models' performance for the entire growing season. In Figure \ref{fig:model_perform_a}, the trajectories of actual states and predicted states of all the schemes are shown for a few selected states. Similarly, as shown in Figure \ref{fig:EKF_perform_a}, the actual and estimated state trajectories of all the schemes are shown for the same states. 

\subsection{Estimation accuracy}
In Figure \ref{fig:scheme1}, Scheme I, the proposed approach, employs specific thresholds for the performance-triggered criterion: $th_e = 40$ for the error threshold and $th_C = 1.0$ for the cluster threshold. The tunable parameter, $\dot{e}_L$, is subject to field-specific variations, with its maximum allowable threshold set at $0.05$. Two different horizon lengths of $N_{f_d}=250$ and $N_{f_d}=350$ are utilized in this scheme. The bottom plots in Figure \ref{fig:scheme1} show the time instances of when the reduced-order model is re-identified and give the dimensionality of the corresponding model. The top plots show the variation of \% MAE  for the reduced model. Model re-identification is conducted when the error indicator $e_L$ exceeds the threshold $th_e$. Initially, at the start of the season, the reduced order model indicates a low model order, indicating a relatively uniform field condition. However, as time progresses and the system receives input, the model order gradually increases, suggesting a shift in field dynamics. Figure \ref{fig:scheme1}(a) showcases an additional model identification due to considerations of a short horizon. Notably, the highest model order is observed for the longer horizon at $2778$, while the short horizon exhibits a model order of $2188$. Figure \ref{surface_map} presents the surface soil pressure head distribution of the actual, estimated, and absolute error between the actual and estimated states for both the $1^{st}$ and $30^{th}$ days, considering the longer horizon. From Figure \ref{surface_map}(a) absolute error plot, it is evident that the absolute estimation error on the first day is significantly smaller than that of the last day, as shown in Figure \ref{surface_map}(b). Similarly, in Figure \ref{rootzone_map}, the soil moisture maps at a depth of $0.3$ \ m are presented, and the absolute error between actual and estimated states is decreased over the estimation period except for a few locations. 

In the non-adaptive model reduction approach, Scheme II, the cluster generation threshold $th_{C}$ is set to $5.0$, resulting in a model order of $1974$ which can capture the dynamics for the total simulation time. As shown in Figure \ref{fig:scheme2_a} the solid green line, the estimation barely converges, with a long convergence delay leading to significant estimation errors. From a practical point of view, the information of all the inputs including weather forecasts will not be perfectly known. In this set of simulations, we consider input disturbances, the input profile of which is shown in the bottom plot of Figure \ref{fig:input}. Using this input information, the estimation performance based on the non-adaptive reduced-order model is relatively sensitive to the disturbances. Specifically, it is observed that the estimation diverges, provided as in the green dotted line in Figure \ref{fig:scheme2_a}. To improve this, we propose a time-triggered adaptive model reduction that can be used for comparison.


Scheme III, which demonstrates time-triggered model reduction, is illustrated in Figure \ref{fig:scheme3} for two different values of $N_{f_d}$. To mitigate the influence of weather forecast and irrigation decision uncertainties, we explore the time-triggered algorithm with structural similarities to Scheme II. In this design, the generated trajectories are updated at shorter fixed intervals, differing from Scheme II, which takes into account the entire operational dataset. The performance of the time-triggered estimator is depicted in red in Figure \ref{fig:scheme2_a}. For $N_{f_d}=250$ and $N_{f_d}=350$, the highest model orders are 2528 and 2868, respectively. When comparing the two triggered EKF designs, performance-triggered and time-triggered, the estimation error remains similar for a specific horizon $N_{f_d}$ considered; however, the time-triggered model reduction exhibits a higher number of model changes. To investigate the plotted results, Figure \ref{fig:scheme2_a} displays how Scheme I with $N_{f_d}=250$ achieves a reduction in estimation error around the $22^{nd}$ day, showcasing the benefits of model reduction.

In Figure \ref{fig:EKF_perform_a}, the state estimate trajectories for all schemes align with the trajectories of the actual state of the system. Figure \ref{fig:scheme2_a} further highlights the percentage of MAE between the actual and EKF estimated states for all schemes. Note that the proposed triggered methods converge much faster compared with the non-adaptive reduced-order model due to the increased degree of observability of the estimation problem. In the proposed triggered schemes, the number of measurements remains the same. Meanwhile, the number of states that need to be estimated is significantly less. This facilitates the convergence of the estimation scheme. However, in the proposed approach, since each reduced model uses fewer nodes and the model mismatch error accumulates, the accuracy after the convergence is slightly poorer as can be seen from Figure ~\ref{fig:scheme2_a}. With the adoption of the proposed model reduction scheme, Scheme I effectively confines the estimation error within the predefined threshold. The tuning parameters for re-identification of the reduced models include the fixed time $N_{f_d}$, threshold for cluster generation $th_{C}$, and error threshold $th_e$. These also provide more flexibility in tuning the performance of the adaptive estimator in the proposed approach.

\subsection{Computation times}
The proposed approach in this study offers significant improvements in computational efficiency. All the computational simulations are performed on a computer loaded with Intel(R) Core(TM) $i7-8700$ CPU operating at $3.2$GHz and $24.0$ GB RAM. In \cite{sarupa_2023}, the algorithm was implemented to a small demo farm of radius $50$\ m and the results were compared against a non-adaptive  reduced model and a full-order state estimator using Richards equation. The findings of the study indicate that the adaptive EKF outperforms the full-order EKF using the Richards equation in terms of estimation accuracy and computational efficiency. By utilizing a smaller field with a total of $9600$ nodes, it becomes more convenient to simulate the EKF for the Richards equation. It is worth noting that with an increasing number of nodes, the computational expenses for solving the Richards equation experience substantial growth. 

\renewcommand{\arraystretch}{1.6}
\begin{table}[t]
	\centering
\begin{tabular}{ |c|c|}  
 \hline
 Scheme & Simulation time (Sec) \\ 
 \hline 
 Scheme I $N_{f_d}=250$ &  5.01\\
 \hline
  Scheme I $N_{f_d}=350$ & 4.93 \\
  \hline
   Scheme II &  2.20\\
   \hline
    Scheme III $N_{f_d}=250$ &  5.52\\
    \hline
     Scheme III $N_{f_d}=350$ &  5.48\\
 \hline
\end{tabular}
\caption{Computational speed per iteration comparison of different schemes}
\label{tbl:speed}
\end{table}

The proposed performance-triggered adaptive EKF, Scheme I, which incorporates adaptive clustering, model reduction, and recursive calculation, requires approximately $5$ seconds to evaluate at each sampling time in Table \ref{tbl:speed}. On the other hand, the time-triggered EKF Scheme III, which experiences frequent model changes, takes a slightly longer estimation time of around $5.5$ seconds per sampling time considering the total simulation time. Compared to Schemes I and III, the non-adaptive reduced model requires less time, although it sacrifices accuracy in the estimation and is prone to input disturbances. If Scheme II results in a high-order reduced model, which may be necessary to accommodate the dynamics over a long period, the evaluation time for calculating the large state transition matrix $A_d$ also increases substantially.


\section{Concluding remarks}
The article addressed the problem of state estimation for a large-scale agricultural field managed with a central pivot. A finite difference method was utilized to discretize the Richards equation, which describes the dynamics of the system within a cylindrical coordinate framework. We designed a reduced state estimator based on a performance-triggered model order reduction method approach. To validate the effectiveness of our approach, we applied the proposed adaptive state estimation to a simulated large-scale agricultural field. The reduced state estimator showcased a satisfactory performance. Improved estimates of the soil moisture were obtained as compared to the non-adaptive reduced-order model. Additionally, improved computational efficiency for large-scale agro-hydrological systems was achieved by using the proposed model order reduction-based approach.

\section{Acknowledgment}
Financial support from Natural Sciences and Engineering Research Council of Canada and Alberta Innovates Technology Futures is gratefully acknowledged. X. Yin acknowledges the support from Ministry of Education, Singapore, under its Academic Research Fund Tier 1 (RG63/22).


\begin{thebibliography}{10}

\bibitem{global_2015}
``Global risks 2015,'' {\em World Economic Forum}, 2015.

\bibitem{wastewater_2017}
``Waste water the untapped resource,'' {\em The United Nations World Water Development Report}, 2017.

\bibitem{lozoya_model_2014}
C.~Lozoya, C.~Mendoza, L.~Mej{\'\i}a, J.~Quintana, G.~Mendoza, M.~Bustillos, O.~Arras, and L.~Sol{\'\i}s, ``Model predictive control for closed-loop irrigation,'' {\em IFAC Proceedings Volumes}, vol.~47, no.~3, pp.~4429--4434, 2014.

\bibitem{goodchild2015method}
M.~S. Goodchild, K.~K{\"u}hn, M.~Jenkins, K.~Burek, and A.~Button, ``A method for precision closed-loop irrigation using a modified pid control algorithm,'' {\em Sensors \& Transducers}, vol.~188, no.~5, p.~61, 2015.

\bibitem{mccarthy2014simulation}
A.~C. McCarthy, N.~H. Hancock, and S.~R. Raine, ``Simulation of irrigation control strategies for cotton using model predictive control within the variwise simulation framework,'' {\em Computers and electronics in agriculture}, vol.~101, pp.~135--147, 2014.

\bibitem{mao2018soil}
Y.~Mao, S.~Liu, J.~Nahar, J.~Liu, and F.~Ding, ``Soil moisture regulation of agro-hydrological systems using zone model predictive control,'' {\em Computers and Electronics in Agriculture}, vol.~154, pp.~239--247, 2018.

\bibitem{bwambale2022smart}
E.~Bwambale, F.~K. Abagale, and G.~K. Anornu, ``Smart irrigation monitoring and control strategies for improving water use efficiency in precision agriculture: A review,'' {\em Agricultural Water Management}, vol.~260, p.~107324, 2022.

\bibitem{saavoss2016yield}
M.~Saavoss, J.~Majsztrik, B.~Belayneh, J.~Lea-Cox, and E.~Lichtenberg, ``Yield, quality and profitability of sensor-controlled irrigation: A case study of snapdragon (antirrhinum majus l.) production,'' {\em Irrigation science}, vol.~34, pp.~409--420, 2016.

\bibitem{lichtenberg2013profitability}
E.~Lichtenberg, J.~Majsztrik, and M.~Saavoss, ``Profitability of sensor-based irrigation in greenhouse and nursery crops,'' {\em HortTechnology}, vol.~23, no.~6, pp.~770--774, 2013.

\bibitem{chappell2013implementation}
M.~Chappell, S.~K. Dove, M.~W. van Iersel, P.~A. Thomas, and J.~Ruter, ``Implementation of wireless sensor networks for irrigation control in three container nurseries,'' {\em HortTechnology}, vol.~23, no.~6, pp.~747--753, 2013.

\bibitem{lu2011dual}
H.~L{\"u}, Z.~Yu, Y.~Zhu, S.~Drake, Z.~Hao, and E.~A. Sudicky, ``Dual state-parameter estimation of root zone soil moisture by optimal parameter estimation and extended kalman filter data assimilation,'' {\em Advances in water resources}, vol.~34, no.~3, pp.~395--406, 2011.

\bibitem{chen2015comparison}
W.~Chen, C.~Huang, H.~Shen, and X.~Li, ``Comparison of ensemble-based state and parameter estimation methods for soil moisture data assimilation,'' {\em Advances in water resources}, vol.~86, pp.~425--438, 2015.

\bibitem{montzka2011hydraulic}
C.~Montzka, H.~Moradkhani, L.~Weiherm{\"u}ller, H.-J.~H. Franssen, M.~Canty, and H.~Vereecken, ``Hydraulic parameter estimation by remotely-sensed top soil moisture observations with the particle filter,'' {\em Journal of hydrology}, vol.~399, no.~3-4, pp.~410--421, 2011.

\bibitem{bo2020parameter}
S.~Bo, S.~R. Sahoo, X.~Yin, J.~Liu, and S.~L. Shah, ``Parameter and state estimation of one-dimensional infiltration processes: A simultaneous approach,'' {\em Mathematics}, vol.~8, no.~1, p.~134, 2020.

\bibitem{yin2021consensus}
X.~Yin, S.~Bo, J.~Liu, and B.~Huang, ``Consensus-based approach for parameter and state estimation of agro-hydrological systems,'' {\em AIChE Journal}, vol.~67, no.~2, p.~e17096, 2021.

\bibitem{agyeman2021soil}
B.~T. Agyeman, S.~Bo, S.~R. Sahoo, X.~Yin, J.~Liu, and S.~L. Shah, ``Soil moisture map construction by sequential data assimilation using an extended kalman filter,'' {\em Journal of Hydrology}, vol.~598, p.~126425, 2021.

\bibitem{bo2020decentralized}
S.~Bo and J.~Liu, ``A decentralized framework for parameter and state estimation of infiltration processes,'' {\em Mathematics}, vol.~8, no.~5, p.~681, 2020.

\bibitem{richards_capillary_1931}
L.~A. Richards, ``Capillary conduction of liquids through porous mediums,'' {\em Physics}, vol.~1, no.~5, pp.~318--333, 1931.

\bibitem{yin2018state}
X.~Yin and J.~Liu, ``State estimation of wastewater treatment plants based on model approximation,'' {\em Computers \& Chemical Engineering}, vol.~111, pp.~79--91, 2018.

\bibitem{kung1981optimal}
S.~Kung and D.~Lin, ``Optimal hankel-norm model reductions: Multivariable systems,'' {\em IEEE Transactions on Automatic Control}, vol.~26, no.~4, pp.~832--852, 1981.

\bibitem{antoulas2005approximation}
A.~C. Antoulas, {\em Approximation of large-scale dynamical systems}.
\newblock SIAM, 2005.

\bibitem{gugercin2004survey}
S.~Gugercin and A.~C. Antoulas, ``A survey of model reduction by balanced truncation and some new results,'' {\em International Journal of Control}, vol.~77, no.~8, pp.~748--766, 2004.

\bibitem{cheng2019gramian}
X.~Cheng and J.~Scherpen, ``Gramian-based model reduction of directed networks,'' {\em arXiv preprint arXiv:1901.01285}, 2019.

\bibitem{sahoo2022knowledge}
S.~R. Sahoo, B.~T. Agyeman, S.~Debnath, and J.~Liu, ``Knowledge-based optimal irrigation scheduling of agro-hydrological systems,'' {\em Sustainability}, vol.~14, no.~3, p.~1304, 2022.

\bibitem{sahoo2022adaptive}
S.~R. Sahoo and J.~Liu, ``Adaptive model reduction and state estimation of agro-hydrological systems,'' {\em Computers and Electronics in Agriculture}, vol.~195, p.~106825, 2022.

\bibitem{rasheed2022soil}
M.~W. Rasheed, J.~Tang, A.~Sarwar, S.~Shah, N.~Saddique, M.~U. Khan, M.~Imran~Khan, S.~Nawaz, R.~R. Shamshiri, M.~Aziz, {\em et~al.}, ``Soil moisture measuring techniques and factors affecting the moisture dynamics: A comprehensive review,'' {\em Sustainability}, vol.~14, no.~18, p.~11538, 2022.

\bibitem{alanqar2017error}
A.~Alanqar, H.~Durand, and P.~D. Christofides, ``Error-triggered online model identification for model-based feedback control,'' {\em AIChE Journal}, vol.~63, no.~3, pp.~949--966, 2017.

\bibitem{steinbach_comparison_2000}
M.~Steinbach, G.~Karypis, and V.~Kumar, ``A comparison of document clustering techniques,'' {\em CS\&E Technical Reports}, 2000.

\bibitem{debnath2022subsystem}
S.~Debnath, S.~R. Sahoo, B.~Decardi-Nelson, and J.~Liu, ``Subsystem decomposition and distributed state estimation of nonlinear processes with implicit time-scale multiplicity,'' {\em AIChE Journal}, vol.~68, no.~5, p.~e17661, 2022.

\bibitem{agyeman2022simultaneous}
B.~T. Agyeman, E.~Orouskhani, and J.~Liu, ``Simultaneous estimation of soil moisture and hydraulic parameters for precision agriculture. part b: Application to a real field,'' in {\em 2022 IEEE International Symposium on Advanced Control of Industrial Processes (AdCONIP)}, pp.~18--23, IEEE, 2022.

\bibitem{sahoo_optimal_2019}
S.~R. Sahoo, X.~Yin, and J.~Liu, ``Optimal sensor placement for agro-hydrological systems,'' {\em AIChE Journal}, vol.~65, no.~12, p.~e16795, 2019.

\bibitem{sarupa_2023}
S.~Debnath, S.~R. Sahoo, B.~T. Agyeman, X.~Yin, and J.~Liu, ``An error-triggered adaptive model reduction and soil moisture estimation for agro-hydrological systems \textit{Accepted},'' in {\em 2023 IEEE Conference on Decision and Control (CDC)}, IEEE, 2023.

\end{thebibliography}
\end{document}